\documentclass{cupconf}
\usepackage{epsfig}
\usepackage{subfigure}
  \checkfont{eurm10}
  \iffontfound
    \IfFileExists{upmath.sty}
      {\typeout{^^JFound AMS Euler Roman fonts on the system,
                   using the 'upmath' package.^^J}%
       \usepackage{upmath}}
      {\typeout{^^JFound AMS Euler Roman fonts on the system, but you
                   dont seem to have the}%
       \typeout{'upmath' package installed. cupconf.cls can take advantage
                 of these fonts,^^Jif you use 'upmath' package.^^J}%
      }
  \else
  \fi

  \checkfont{msam10}
  \iffontfound
    \IfFileExists{amssymb.sty}
      {\typeout{^^JFound AMS Symbol fonts on the system, using the
                'amssymb' package.^^J}%
       \usepackage{amssymb}%
       \let\le=\leqslant  
         
      }{}
  \fi

  \IfFileExists{amsbsy.sty}
    {\typeout{^^JFound the 'amsbsy' package on the system, using it.^^J}%
     \usepackage{amsbsy}}
    {}

\newsavebox{\astrutbox}
\sbox{\astrutbox}{\rule[-5pt]{0pt}{20pt}}

\title[Dynamics around supermassive black holes]
{Dynamics around supermassive black holes}

\author[A. Gualandris \& D. Merritt]{%
A\ls L\ls E\ls S\ls S\ls I\ls A\ls \ns G\ls U\ls A\ls L\ls A\ls
N\ls D\ls R\ls I\ls S\ls \ns \and\ns
D\ls A\ls V\ls I\ls D\ls \ns M\ls E\ls R\ls R\ls I\ls T\ls T\ls}

\affiliation{Department of Physics and Center for Computational 
Relativity and Gravitation, Rochester Institute of Technology,
Rochester, USA}

\pubyear{2007}
\volume{000}
\pagerange{000--000}
\date{?? and in revised form ??}
\begin{document}

\maketitle

\begin{abstract}
The dynamics of galactic nuclei reflects the presence of supermassive 
black holes (SBHs) in many ways.  
Single SBHs act as sinks, destroying a mass 
in stars equal to their own mass in roughly one relaxation time and 
forcing nuclei to expand. 
Formation of binary SBHs displaces a mass in stars roughly equal to
the binary mass, creating low-density cores and 
ejecting hyper-velocity stars.  
Gravitational radiation recoil can eject coalescing binary SBHs 
from nuclei, resulting in offset SBHs and lopsided cores.
We review recent work on these mechanisms and discuss the
observable consequences.
\end{abstract}

\firstsection % if your document starts with a section,
\section{Characteristic scales}

Supermassive black holes (SBHs) are ubiquitous components of
bright galaxies and many have been present
with roughly their current masses
($\sim 10^9 M_\odot$) since very early times, as soon as 
$\sim 1$ Gyr after the Big Bang
(\cite[Fan {\it et~al.} 2003]{Fan:03};
\cite[Marconi {\it et~al.} 2004]{Marconi:04}).
A SBH strongly influences the motion of stars within a
distance $r_h$, the gravitational influence radius, where
\begin{equation}
r_h = {GM_\bullet\over \sigma_c^2};
\end{equation}
$M_\bullet$ is the SBH mass and $\sigma_c$ is the stellar
(1d) velocity dispersion in the core.
Using the tight empirical
correlation between $M_\bullet$ and $\sigma_c$:
\begin{equation}
\left({M_\bullet\over 10^8M_\odot}\right) = (1.66\pm 0.24) 
\left({\sigma_c\over 200\ {\rm km\ s}^{-1}}\right)^\alpha, \ \ \ 
\alpha = 4.86\pm 0.4
\label{eq:ms}
\end{equation}
(\cite[Ferrarese \& Ford 2005]{FF:05}),
this can be written
\begin{equation}
r_h \approx 18\ {\rm pc} \left({\sigma_c\over 200\ 
{\rm km\ s}^{-1}}\right)^{2.86} 
\approx 13\ {\rm pc} \left({M_\bullet\over 10^8M_\odot}\right)^{0.59}.
\label{eq:rhnew} 
\end{equation}
While the velocities of stars must increase -- by definition --
inside $r_h$, this radius is not
necessarily associated with any other observational marker.
Such is the case at the Galactic center, for instance, where the stellar
density exhibits no obvious feature at $r_h\approx 3$ pc.
However the most luminous elliptical galaxies always
have cores, regions near the center where the stellar density
is relatively low.
Core radii are of order $r_h$ in these galaxies,
and the stellar mass that was (apparently) removed in
creating the core is of order $M_\bullet$.
These facts suggest a connection between the cores and
the SBHs, and this idea has motivated much recent work,
reviewed here, on binary SBHs and on the consequences of displacing SBHs
temporarily or permanently from their central locations in galaxies.

\begin{figure}
  \vspace{0.5cm}
  \includegraphics[height=0.35\textheight]{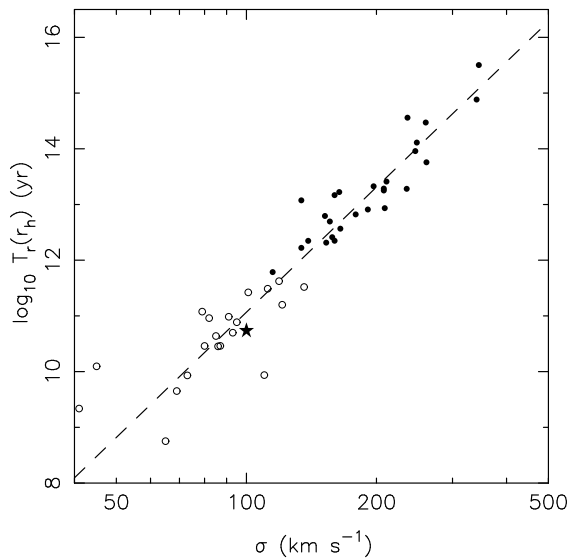}
  \caption{Relaxation time, measured at the SBH influence radius,
   in a sample of early-type galaxies (C\^ot\'e et al. 2004), {\it vs}.
   the central stellar velocity dispersion. Filled symbols are
   galaxies in which the SBH's influence radius is resolved; the
   star is the Milky Way bulge.
   (From Merritt, Mikkola \& Szell 2007).}\label{fig:trrh}
\end{figure}

An important time scale associated with galactic nuclei
(not just those containing SBHs) is the relaxation time, defined as 
the time for 
gravitational encounters between stars to establish a locally 
Maxwellian velocity distribution.
The nuclear relaxation time is
(\cite[Spitzer 1987]{Spitzer:87})
\begin{subequations}
\begin{eqnarray}
T_R &\approx& {0.34\sigma_c^3\over G^2 \rho_c m_\star \ln\Lambda} \\
&\approx& 1.2\times 10^{10}\ {\rm yr}
\left({\sigma_c\over 100\ {\rm km\ s}^{-1}}\right)^3
\left({\rho_c\over 10^5M_\odot {\rm pc}^{-3}}\right)^{-1} 
\left({m_\star\over M_\odot}\right)^{-1} 
\left({\ln\Lambda\over 15}\right)^{-1}
\label{eq:tr}
\end{eqnarray}
\end{subequations}
with $\rho_c$ the nuclear density 
and $\ln\Lambda$ the Coulomb logarithm.
Figure~\ref{fig:trrh} shows estimates of $T_R$
measured at $r_h$ in a sample of early-type galaxies,
assuming $m_\star = 1M_\odot$.
A least-squares fit to the points (shown as the dashed line
in the figure) gives
\begin{equation}
T_R(r_h) \approx 2.5\times 10^{13}\ {\rm yr}
\left({\sigma\over 200\ {\rm km\ s}^{-1}}\right)^{7.47}
\approx 9.6\times 10^{12}\ {\rm yr}
\left({M_\bullet\over 10^8 M_\odot}\right)^{1.54}.
\label{eq:combine}
\end{equation}
``Collisional'' nuclei can be defined as those with
$T_R(r_h)\lesssim 10$ Gyr; figure~\ref{fig:trrh} shows
that such nuclei are uniquely associated with galaxies that 
are relatively faint,
as faint as or fainter than the Milky Way bulge,
which has $T_R(r_h)\approx 4\times 10^{10}$ yr.
Furthermore, relaxation-driven changes in the stellar distribution
around a SBH are generally confined to radii $\lesssim 10^{-1} r_h$,
making them all but unobservable in galaxies beyond the 
Local Group (T. Alexander, these proceedings).
But the relaxation time also fixes the rate of
gravitational scattering of stars into the central ``sink'' -- 
either a single or a binary SBH -- and this fact has important
consequences for nuclear evolution in low-luminosity galaxies, 
as discussed below.

\section{Core structure}\label{sec:corestructure}

The `core' of a galaxy can loosely be defined as the region
near the center where the density of starlight drops
significantly below what is expected based on an inward extrapolation
of the overall luminosity profile.
At large radii, the surface brightness profiles of early-type 
galaxies are well fit by the \cite{Sersic:68} model,
\begin{equation}
\ln I(R) = \ln I_e - b(n)\left[\left(R/R_e\right)^{1/n} - 1\right].
\label{eq:sersic}
\end{equation}
The quantity $b$ is normally chosen such that $R_e$ is the projected
radius containing one-half of the total light.
The shape of the profile is then determined by $n$;
$n= 4$ is the \cite{DeVauc:48} model, which is a good
representation of bright elliptical (E) galaxies
(\cite[Kormendy \& Djorgovski 1989]{DK:89}),
while $n=1$ is the exponential model, which approximates
the luminosity profiles of dwarf elliptical (dE) galaxies 
(\cite[Binggeli, Sandage \& Tarenghi 1984]{Binggeli:84}).
An alternative way to write (\ref{eq:sersic}) is
\begin{equation}
{d\ln I\over d\ln R} = -{b\over n}\left({R\over R_e}\right)^{1/n},
\label{eq:serslopes}
\end{equation}
i.e. the logarithmic slope varies as a power of the projected radius.
While there is no consensus on why the
S\'ersic model is such a good representation
of stellar spheroids, a possible hint comes from the
dark-matter halos produced in 
$N$-body simulations of hierarchical structure
formation, which are also well described by (\ref{eq:serslopes})
(\cite[Navarro {\it et~al.} 2004]{Navarro:04}),
suggesting that S\'ersic's model applies generally to systems that form
via dissipationless clustering
(\cite[Merritt {\it et~al.} 2005]{Universal:05}).

S\'ersic's model is known to accurately reproduce the luminosity
profiles in some galaxies over at least three decades
in radius (e.g. \cite[Graham {\it et~al.} 2003]{Graham:03}), 
but deviations often appear near the center.
Galaxies fainter than absolute magnitude 
$M_B\approx -19$ tend to have {\it higher} central surface brightness
than predicted by S\'ersic's model;
the structure of the central excess is typically unresolved but its
properties are consistent with those of a compact, 
intermediate-age star cluster
(\cite[Carollo, Stiavelli \& Mack 1998]{Carollo:98};
\cite[C\^ot\'e {\it et~al.} 2006]{ACS8};
\cite[Balcells {\it et~al.} 2007]{Balcells:07}).
Galaxies brighter than $M_B\approx -20$ have long been known
to exhibit central {\it deficits}
(e.g. \cite[Kormendy 1985a]{Kormendy:85a});
these have traditionally been called simply ``cores,'' 
perhaps because a flat central density profile was considered 
a priori most natural (\cite[Tremaine 1997]{Tremaine:97}).

For about two decades, it was widely  believed that
dE galaxies were distinct objects
from the more luminous E galaxies.
The dividing line between the two classes was put 
at absolute magnitude $M_B\approx -18$,
based partly on the presence of cores in bright galaxies,
and also on the relation between total luminosity and mean
surface brightness
(\cite[Kormendy 1985b]{Kormendy:85b}).
This view was challenged by \cite{Jerjen:97},
and in a compelling series of papers, A. Graham and
collaborators showed that -- aside from the cores --
early-type galaxies display a remarkable continuity of 
structural properties,
from $M_B\approx -13$ to $M_B\approx -22$
(\cite[Graham \& Guzman 2003]{GG:03}; 
\cite[Graham {\it et~al.} 2003]{Graham:03}; 
\cite[Trujillo {\it et~al.} 2004]{Trujillo:04}). 

\begin{figure}[t]
\begin{center}
\includegraphics[height=0.30\textheight]{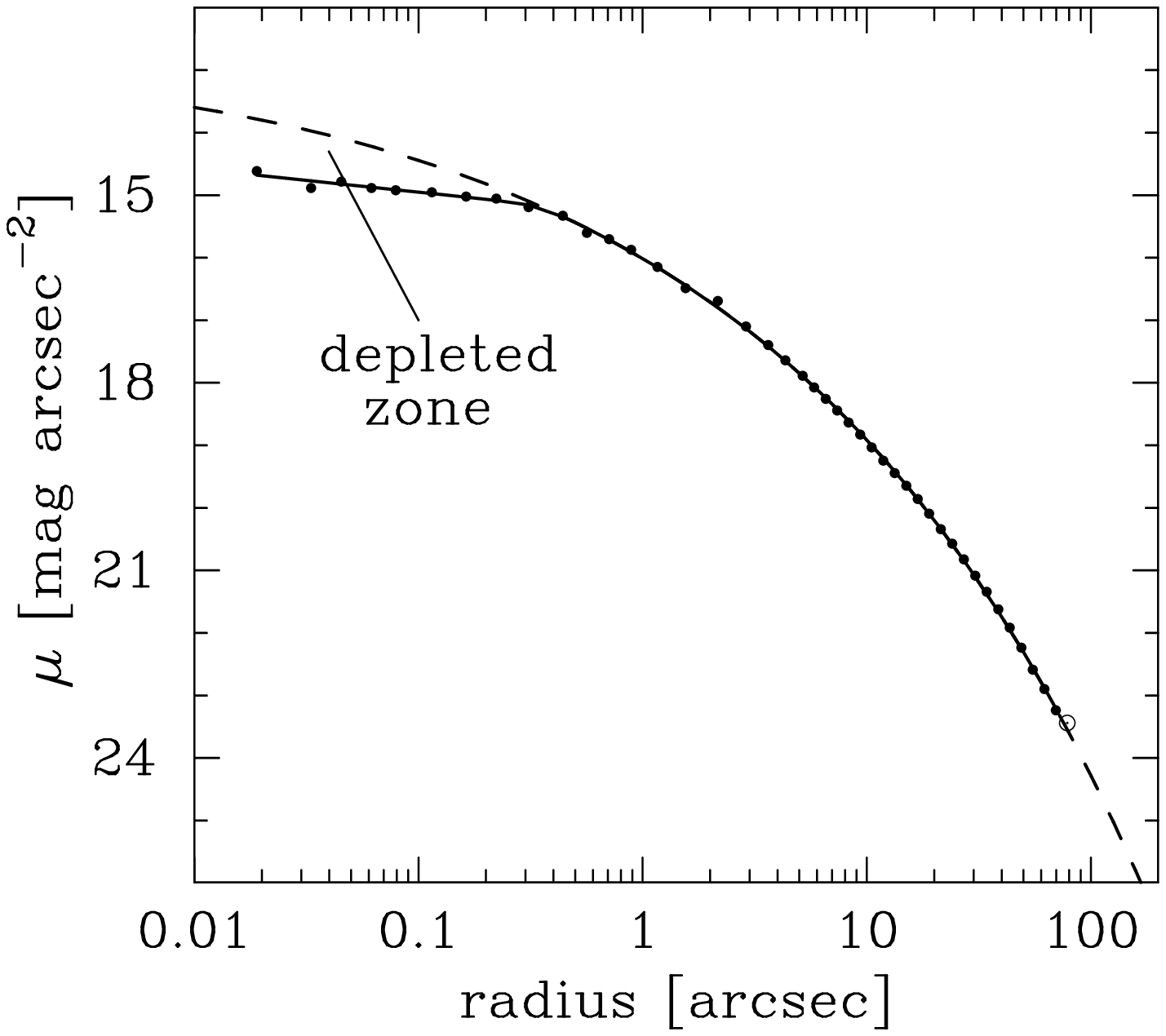}
\includegraphics[height=0.30\textheight]{fig_mdefhist.ps}
\end{center}
\caption[]{{\it Left:} Surface brightness profile in the $R$ band
of NGC 3348, a ``core'' galaxy.
The dashed line is the best-fitting S\'ersic model;
the observed profile (points, and solid line) falls
below this inside of a break radius $r_b\approx 0''.35$.
(From Graham 2004.)
{\it Right:} Histogram of observed mass deficits for the
sample of core galaxies in Graham (2004) 
and Ferrarese {\it et al.} (2006).
(Adapted from Merritt 2006a.)}
\label{fig:mdef}
\end{figure}

The connection between nuclear star clusters and SBHs,
if any, is unclear; in fact it has been suggested that the two
are mutually exclusive 
(\cite[Ferrarese {\it et~al.} 2006]{Ferrarese:06};
\cite[Wehner \& Harris 2006]{Wehner:06}),
although counter-examples to this rule probably exist,
e.g. NGC 3384 which contains a nuclear cluster
(\cite[Ravindranath {\it et~al.} 2001]{Ravin:01})
and may contain a SBH
(\cite[Gebhardt {\it et~al.} 2003]{Gebhardt:03}).

Here we focus on the cores.
The cores extend outward to a break radius $r_b$ 
that is roughly a few times $r_h$,
or from $\sim 0.01$ to $\sim 0.05$ times $R_e$.
A more robust way of quantifying the cores is in
terms of their mass (i.e. light): the ``mass deficit''
(\cite[Milosavljevi{\'c} {\it et~al.} 2002]{Milos:02})
is defined as the difference in integrated mass 
between the observed density profile $\rho(r)$ and an
inward extrapolation of the outer profile,
$\rho_{out}(r)$, typically modelled as a S\'ersic profile
(figure~\ref{fig:mdef}):
\begin{equation}
M_{def}\equiv 4\pi \int_0^{r_b} 
\left[\rho_{out}(r) -\rho(r)\right] r^2 dr .
\label{eq:defmdef}
\end{equation}
Figure~\ref{fig:mdef} shows mass deficits for a sample of
``core'' galaxies, expressed in units of the SBH mass.
There is a clear peak at $M_{\rm def}\approx 1 M_\bullet$,
although some galaxies have much larger cores.

The fact that core and SBH masses are often so similar 
suggests a connection between the two.
Ejection of stars by binary SBHs during galaxy mergers
is a natural model (\cite[Begelman, Blandford \& Rees 1980]{BBR:80});
the non-existence of cores in fainter galaxies could
then be due to regeneration of a steeper density profile
by star formation (e.g. \cite[McLaughlin {\it et.~al} 2006]{King:06})
or by dynamical evolution associated with the (relatively) short
relaxation times in faint galaxies 
(e.g. \cite[Merritt \& Szell 2006]{MS:06}).
However the largest cores are difficult to explain
via the binary model (\cite[Milosavljevi{\'c} \& Merritt 2001]{MM:01}).

\section{Massive binaries}\label{sec:massivebinaries}

A typical mass ratio for galaxy mergers in the local Universe
is $\sim 10:1$ (e.g. \cite[Sesana {\it et~al.} 2004]{Sesana:04}).
To a good approximation, the initial approach of the two
SBHs can therefore be modelled by assuming that the galaxy hosting
the smaller BH spirals inward under the influence 
of dynamical friction from the fixed distribution of
stars in the larger galaxy.
Modelling both galaxies as singular isothermal
spheres ($\rho\sim r^{-2}$) and assuming that the smaller
galaxy spirals in on a circular orbit, its tidally-truncated mass
is $m_2\approx \sigma_2^3r/2G\sigma$
where $\sigma_2$ and $\sigma$ are the velocity dispersion of the
small and large galaxy respectively
(\cite[Merritt 1984]{Merritt:84}).
Chandrasekhar's (1943) formula then gives for the orbital
decay rate and infall time 
\begin{equation}
{dr\over dt} = -0.30 {Gm_2\over\sigma r}\ln\Lambda,\ \ \ \ 
t_{infall} \approx 3.3 {r(0)\sigma^2\over \sigma_2^3}
\label{eq:df}
\end{equation}
where $\ln\Lambda$ has been set to 2.
Using (\ref{eq:ms})
to relate $\sigma$ and $\sigma_2$ to the respective
SBH masses $M_1$ and $M_2$, this becomes
\begin{equation}
t_{infall}\approx 3.3{r(0)\over\sigma} 
\left({M_2\over M_1}\right)^{-0.62},
\end{equation}
i.e. $t_{infall}$ exceeds the crossing time of the larger
galaxy by a factor $\sim q^{-0.6}$, $q\equiv M_2/M_1\le 1$.
Thus for mass ratios $q\gtrsim 10^{-3}$, 
infall requires less than $\sim 10^2 T_{cr}\approx 10^{10}$ yr.
This mass ratio is roughly the ratio between the masses of the
largest ($\sim 10^{9.5}M_\odot$) and smallest 
($\sim 10^{6.5}M_\odot$) known SBHs and so it is reasonable
to assume that galaxy mergers will almost always lead to formation of
a binary SBH in a time less than $10$ Gyr.
This conclusion is strengthened if the effects of gas are 
taken into account (e.g. \cite[Mayer {\it et~al.} 2007]{Mayer:07}).

Equation (\ref{eq:df}) begins to break down when the
two SBHs approach more closely than $\sim r_h$,
the influence radius of the larger hole,
since the orbital energy of $M_2$ is absorbed by the stars, 
lowering their density and reducing the frictional force.
In spite of this slowdown, $N$-body integrations 
(\cite[Merritt \& Cruz 2001]{MC:01};
\cite[Merritt \& Milosavljevi{\'c} 2001]{MM:01};
\cite[Makino \& Funato 2004]{Makino:04};
\cite[Berczik {\it et~al.} 2005]{Berczik:05})
show that
the separation between the two SBHs continues to
drop rapidly until the binary semi-major axis
is $a\approx a_h$, where
\begin{subequations}
\begin{eqnarray}
a_h &\equiv& {G\mu\over 4\sigma^2} \approx {1\over 4}{q\over (1+q)^2}r_h \\
&\approx& 3.3 {\rm pc} {q\over (1+q)^2} \left({M_1+M_2\over 10^8 M_\odot}\right)^{0.59}
\end{eqnarray}
\label{eq:ah}
\end{subequations}
and $\mu\equiv M_1M_2/(M_1+M_2)$ is the binary reduced mass.
At this separation -- the ``hard binary'' separation --
the binary's binding energy per unit mass
is $\sim\sigma^2$ and it
ejects stars that pass within a distance $\sim a$
with velocities large enough to remove them from the nucleus 
(\cite[Mikkola \& Valtonen 1992]{MV:92};
\cite[Quinlan 1996]{Quinlan:96}).

\begin{figure}[t]
\vspace{-0.5cm}
\begin{center}
\includegraphics[height=0.60\textheight,angle=-90.]{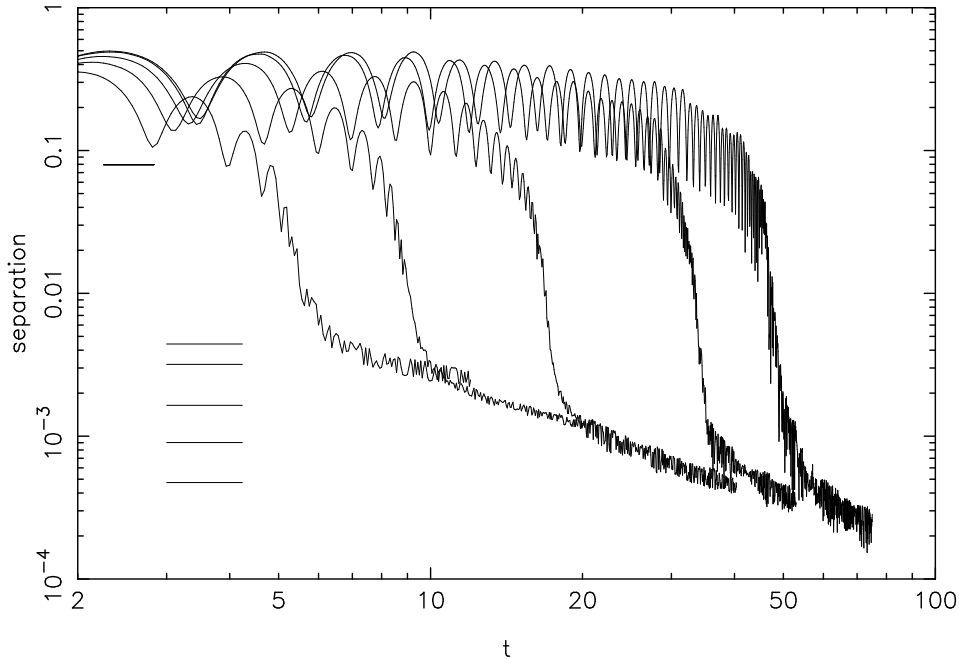}
\end{center}
\caption[]{Early evolution of binary SBHs in $N$-body galaxies, 
for different values of the binary mass ratio,
$M_2/M_2=0.5,0.25,0.1,0.05,0.025$ (left to right).
The upper horizontal line indicates $r_h$,
the influence radius of the more massive hole.
Lower horizontal lines show $a_h$ (\ref{eq:ah}),
the ``hard-binary'' separation.
The evolution of the binary slows drastically when
$a\approx a_h$; in a real (spherical) galaxy with
much larger $N$, evolution would stall at this separation.
The smaller the infalling BH, the farther it spirals in before
stalling.
(Adapted from Merritt 2006a.)}
\label{fig:rbinoft}
\end{figure}

What happens next depends on the density and geometry 
of the nucleus.
In a spherical or axisymmetric galaxy, 
the mass in stars on orbits that intersect the binary
is small, $\lesssim M_1+M_2$, and the 
binary rapidly ejects these stars;
no stars then remain to interact with the binary and its evolution
stalls (figure~\ref{fig:rbinoft}).
In non-axisymmetric (e.g. triaxial) nuclei,
the mass in stars on
centrophilic orbits can be much larger, allowing the binary
to continue shrinking past $a_h$.
And in collisional nuclei of any geometry, gravitational scattering
of stars can repopulate depleted orbits.
These different possbilities are discussed in more detail below.

If the binary does stall at $a\approx a_h$,
it will have given up an energy
\begin{subequations}
\begin{eqnarray}
\Delta E &\approx& -{GM_1M_2\over 2r_h} + {GM_1M_2\over 2a_h} \\
&\approx& -{1\over 2}M_2\sigma^2 + 2(M_1+M_2)\sigma^2 \\
&\approx& 2 (M_1+M_2)\sigma^2
\end{eqnarray}
\end{subequations}
to the stars in the nucleus, i.e.,
the energy transferred from the binary to 
the stars is roughly proportional to the 
{\it combined} mass of the two SBHs.
The reason for this counter-intuitive 
result is the $a_h\propto M_2$
dependence of the stalling radius
(\ref{eq:ah}): smaller infalling BHs
form tighter binaries.
Detailed $N$-body simulations (\cite[Merritt 2006a]{Merritt:06}) 
verify that the mass deficit generated by the binary in evolving from
$\sim r_h$ to $\sim a_h$ is a weak function of the mass ratio,
\begin{equation}
M_{\rm def,h} \approx 0.4(M_1+M_2) \left({q\over 0.1}\right)^{0.2}
\label{eq:mdefh}
\end{equation}
for $0.025\lesssim q\lesssim 0.5$.
A mass deficit of $\sim 0.5 M_\bullet$ is still a factor 
$\sim 2$ too small to explain the observed peak
in the $M_{\rm def}/M_\bullet$ histogram
(figure~\ref{fig:mdef}).
On the other hand, bright elliptical galaxies 
have probably undergone numerous mergers,
and the proportionality between $M_{\rm def}$ and
$M_1+M_2$ (rather than, say, $M_2$) implies that 
the mass deficit following ${\cal N}$ mergers 
is $\sim 0.5{\cal N}$ times the {\it accumulated} BH mass.
Mass deficits in the range
$0.5\lesssim M_{\rm def}/M_\bullet\lesssim 1.5$
therefore imply $1\lesssim {\cal N}\lesssim 3$ mergers,
consistent with the number of major
mergers expected for bright galaxies since the epooch at which
most of the gas was depleted
(e.g. \cite[Haehnelt \& Kauffmann 2002]{HK:02}).
Hierarchical growth of cores tends to saturate after a few
mergers however 
making it difficult to explain mass deficits greater than
$\sim 2 M_\bullet$ in this way.
An effective way to enlarge cores still more is to kick
the SBH out, at least temporarily, as discussed in \S 4.

The first convincing evidence for a true, binary SBH 
was recently presented by \cite{Rodriguez:06},
who discovered two compact, flat-spectrum AGN 
at the center of a single elliptical galaxy,
with a projected separation of $\sim 7$ pc.
This is consistent with the radius at formation of a
$\sim 10^{7.5} M_\odot$ binary (\ref{eq:rhnew}), or the stalling
radius for a binary of at least $\sim 10^{9.3}M_\odot$ 
(\ref{eq:ah}).
Rodriguez {\it et~al.} estimate a binary mass of 
$\sim 10^8 M_\odot$ but with considerable uncertainty.
All other examples of ``binary'' SBHs in single galaxies
have separations $\gg r_h$ (\cite[Komossa 2006]{Komossa:06}).

\begin{figure}
\vspace{-1.0cm}
\centering
\includegraphics[height=0.65\textheight,angle=-90.]{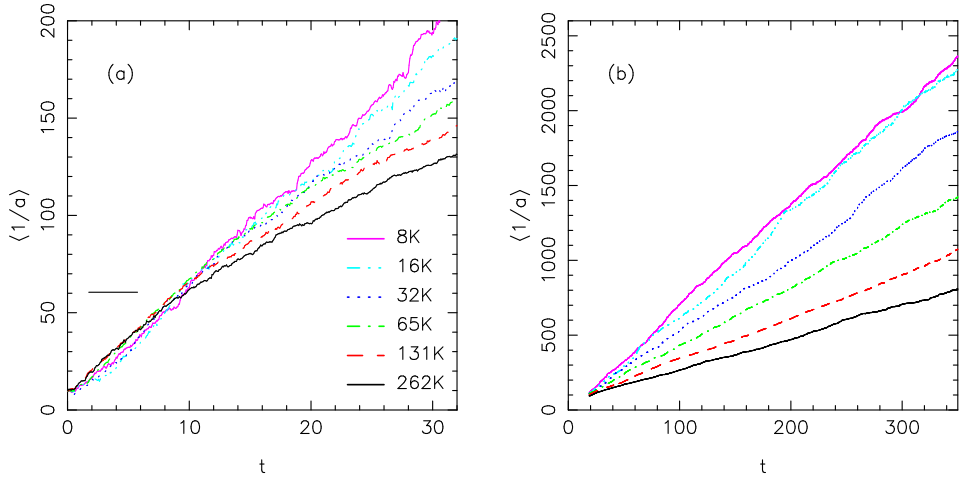}
\caption{Short-term {\it (a)} and long-term {\it (b)}
evolution of a massive binary in a series of $N$-body integrations.
Vertical axis is the inverse semi-major axis (i.e. energy) of the binary,
computed by averaging several independent $N$-body runs;
different curves correspond to different values of $N$, 
the number of ``star'' particles. 
The evolution of the binary is independent of $N$ until 
$a\approx a_h$ (horizontal line); thereafter
the evolution rate is limited by how quickly stars are scattered
onto orbits that intersect the binary, and decreases with
increasing $N$.
(From Merritt, Mikkola \& Szell 2007.)
\label{fig:MMS6}
}
\end{figure}

Even in a spherical galaxy, the stalling that occurs at 
$a\approx a_h$ can be avoided if stars continue to be scattered 
onto orbits that intersect the binary 
(\cite[Valtonen 1996]{Valtonen:96}; \cite[Yu 2002]{Yu:02};
\cite[Milosavljevi{\'c} \& Merritt 2003]{MM:03}).
Such ``collisional loss-cone repopulation'' requires that 
the two-body (star-star) relaxation time at $r\approx r_h$ 
be less then $\sim 10^{10}$ yr; 
according to (\ref{eq:combine}), this is the case in
galaxies with $M_\bullet\lesssim 10^6M_\odot$,
i.e. at the extreme low end of the SBH mass distribution.
%Even the Milky Way nucleus ($T_R(r_h)\approx 10^{10.5}$ yr) 
%is barely in this regime.
Collisional loss cone repopulation is therefore irrelevant to 
the luminous galaxies that are observed to have cores
but may be important in the mass range 
($M_\bullet \lesssim 10^7 M_\odot$) of most interest
to space-based gravitational wave interferometers like LISA
(\cite[Hughes 2006]{Hughes:06}).

\begin{figure}
  \vspace{0.5cm}
  \includegraphics[height=0.36\textheight]{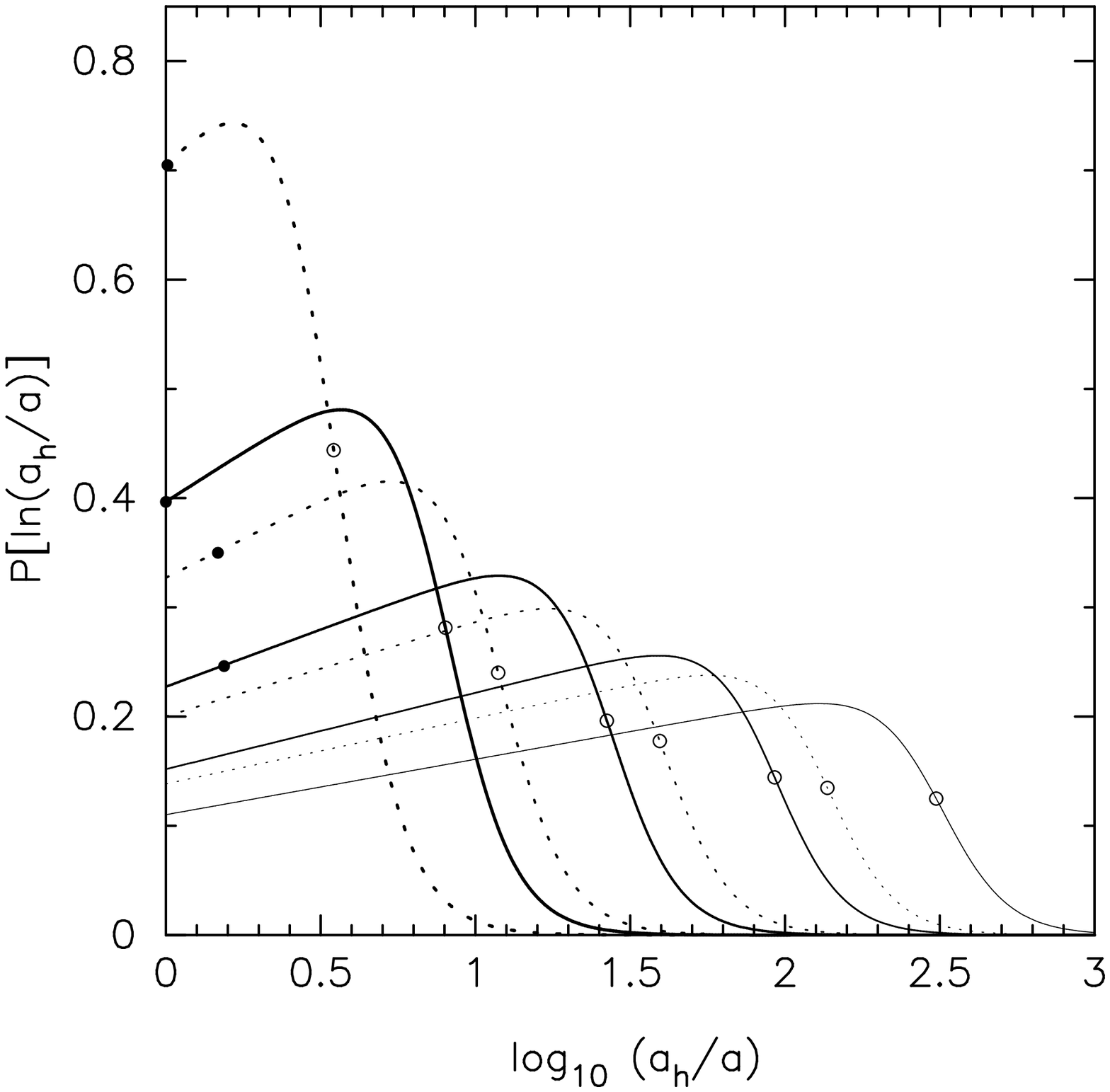}
  \includegraphics[height=0.36\textheight]{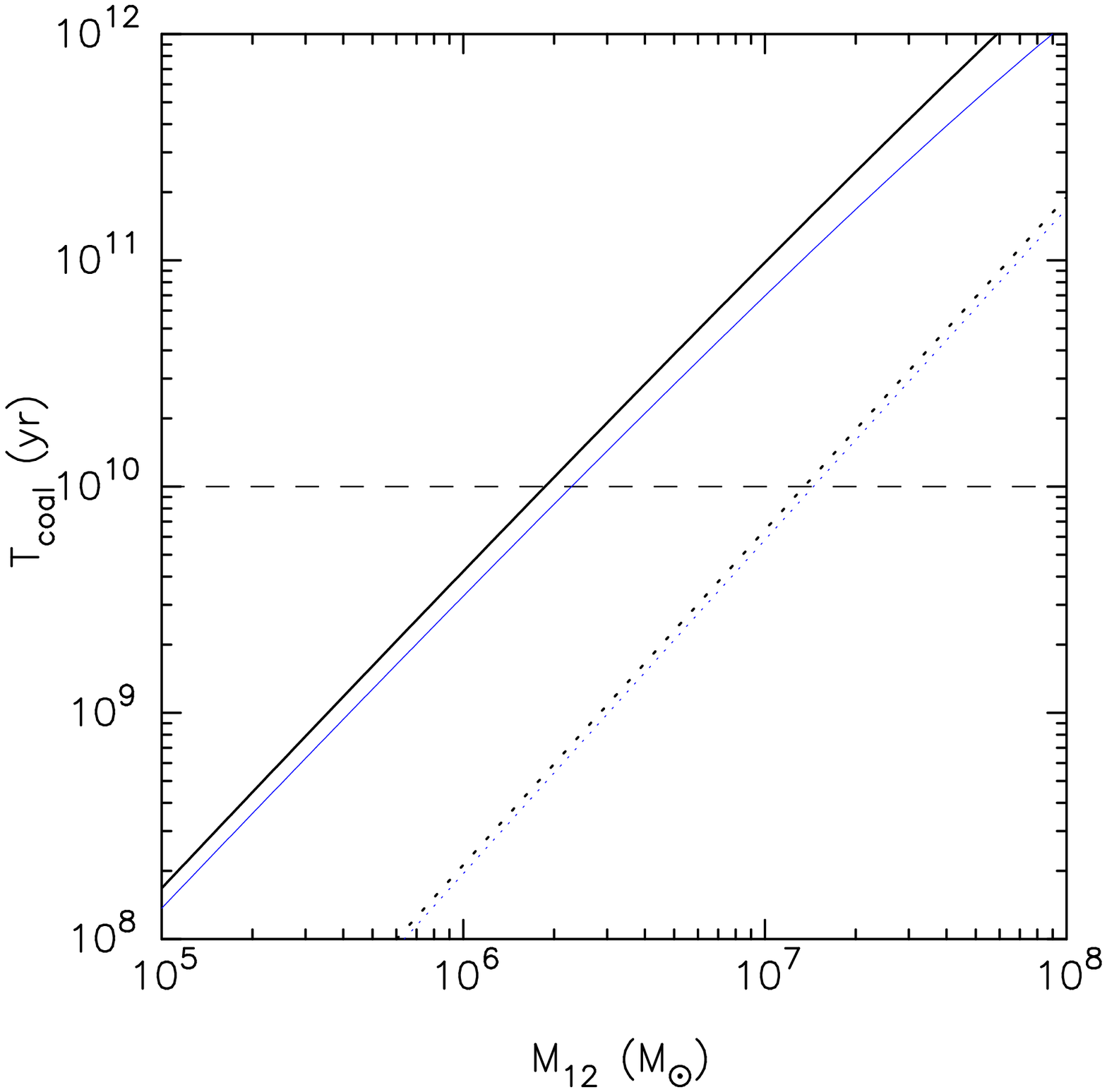}
  \caption{Evolution of a binary SBH in a collisional nucleus,
   based on a Fokker-Planck model that allows for evolution
   of the stellar distribution (Merritt, Mikkola \& Szell 2007).
   {\it Left:} Probability of finding the binary in a unit
   interval of $\ln a$.
   From left to right, curves are for 
   $M_1+M_2=(0.1,1,10,100)\times 10^6M_\odot$.
   Solid (dashed) curves are for $M_2/M_1 =1 (0.1)$.
   Open circles indicate when the rate of energy loss to
   stars equals the loss rate to gravitational waves; 
   filled circles
   correspond to an elapsed time since $a=a_h$ of $10^{10}$ yr.
   For the two smallest values of $M_\bullet$, the latter time
   occurs off the graph to the right.
   {\it Right:} Total time for a binary to evolve from $a=a_h$
   to gravitational wave coalescence as 
   a function of binary mass.
   The thick (black) curve is for $M_2/M_1=1$ and the 
   thin (blue) curve is for $M_2/M_1=0.1$.
   Dotted curves show the time spent in the gravitational
   radiation regime only.
   \label{fig:MMS1920}}
\end{figure}

$N$-body simulation would seem well suited to this problem
(e.g. \cite[Governato, Colpi \& Maraschi 1994]{Governator:94};
\cite[Makino 1997]{Makino:97};
\cite[Milosavljevi{\'c} \& Merritt 2001]{MM:01};
\cite[Makino \& Funato 2004]{Makino:04}).
The difficulty is noise -- or more precisely, getting the level
of noise just right.
In a real galaxy there is a clear separation of time scales
between an orbital period, and the time for stars to be scattered
onto depleted orbits: the first is typically much shorter than the second
which means that orbits intersecting the binary will remain empty 
for many periods before a new star is scattered in.
This is called the ``empty loss cone'' regime and it implies that
supply of stars to the binary will take place diffusively.
In an $N$-body simulation, however, $N$ is much smaller than its value
in real galaxies and orbits are repopulated too quickly.
This is the reason that binary evolution rates in $N$-body simulations
typically scale as $N^{-\alpha}, \alpha<1$ rather than the
$\sim T_R^{-1}\sim N^{-1}$ dependence expected if stars diffused gradually into
an empty loss cone (\cite[Merritt \& Milosavljevi{\'c} 2003]{MM:03}).
Figure~\ref{fig:MMS6} provides an illustration:
early evolution of the binary, until $a\approx a_h$, is $N$-independent;
formation of a hard binary then depletes the loss cone
and continued hardening occurs at a rate that
is a decreasing function of $N$, though less steep than $N^{-1}$.

An alternative approach is based on the Fokker-Planck equation.
Both single and binary SBHs can be modelled as 
``sinks'' located at the centers of galaxies (\cite[Yu 2002]{Yu:02}).
The main differences are the larger physical extent
of the binary ($\sim G(M_1+M_2)/\sigma^2$ vs. $GM_\bullet/c^2$)
and the fact that the binary gives stars a finite kick
rather than disrupting or consuming them completely.
However the diffusion rate of stars into a central sink
varies only logarithmically with the size of the sink
(\cite[Lightman \& Shapiro 1978]{LS:78}),
and a hard binary ejects most stars well out of
the core with $V\gg\sigma$, so the analogy is fairly good.
The Fokker-Planck equation describing nuclei with sinks
is (\cite[Bahcall \& Wolf 1977]{BW:77})
\begin{equation}
{\partial N\over\partial t} = 4\pi^2 p(E) {\partial f\over \partial t}
= -{\partial F_E\over\partial E} - {\cal F}(E,t).
\label{eq:dndt}
\end{equation}
Here $N(E,t)=4\pi p(E)f(E,t)$ is the distribution of stellar energies,
$f(E,t)$ is the phase space density and
$p(E)$ is a phase-space volume element.
The first term on the RHS of (\ref{eq:dndt}) describes the response
of $f$ to the flux $F_E$ of stars in energy space due 
to encounters.
The second term, $-{\cal F}$, is the flux of stars into the
sink, which is dominated by scattering in angular momentum
(\cite[Frank \& Rees 1976]{FR:76}).
A proper treatment of the latter term requires a 2d (energy, angular
momentum) analysis, but a good approximation to ${\cal F}$
can be derived by assuming that the distribution of stars
has reached a quasi steady-state
near the loss cone boundary in phase space
(\cite[Cohn \& Kulsrud 1979]{CK:79}).
If the sink is a binary SBH, a second equation is needed
that relates the flux of stars into the loss cone to
 the rate of change of the binary's semi-major axis:
\begin{equation}
{d\over dt}\left({1\over a}\right) = {2\langle C\rangle\over a(M_1+M_2)}
\int {\cal F}(E,t) dE
\label{eq:hardb}
\end{equation}
with $\langle C\rangle\approx 1.25$ a dimensionless mean energy change for
stars that interact with the binary.

Both terms on the RHS of (\ref{eq:dndt}) imply changes in a time
$\sim T_R$.
The first term on its own implies evolution toward the Bahcall-Wolf (1976)
``zero flux'' solution, $\rho\sim r^{-7/4}$.
The second term implies that a mass of order $\sim M_\bullet$
will be scattered into the sink
in a time of $T_R(r_h)$.
When the sink is a binary SBH, the binary responds by 
ejecting the incoming stars and shrinking,
according to (\ref{eq:hardb}).
As a result, changes in the structure of the nucleus on a relaxation
time scale
(e.g. growth of a core) are 
directly connected to changes in the binary semi-major axis.

Numerical solutions to (\ref{eq:dndt}), (\ref{eq:hardb}) 
(including also the effects of a changing gravitational potential)
have been presented by \cite{MMS:07}.
The solutions are well fit by
\begin{equation}
\ln\left({a_h\over a}\right) = -{B\over A} + \sqrt{{B^2\over A^2}
+ {2\over A}{t\over T_R(r_h)}}
\label{eq:tdepend}
\end{equation}
where $t$ is defined as the time since the binary first became
hard ($a=a_h$), and the coefficients
$A\approx 0.016$, $B\approx 0.08$ depend weakly 
on the binary mass ratio.
Including the effect of energy lost to gravitational radiation:
\begin{equation}
{d\over dt}\left({1\over a}\right) = 
{d\over dt}\left({1\over a}\right)_{\rm stars} +  
{d\over dt}\left({1\over a}\right)_{\rm GR}
\end{equation}
allows one to compute the time to full coalescence,
$T_{\rm coal}$.
Figure~\ref{fig:MMS1920} shows $T_{\rm coal}$ (right panel),
and the time spent by the binary
in unit intervals of $\ln a$ prior to coalescence
(left panel), as functions of binary mass.
The time to coalescence is well fit by
\begin{subequations}
\begin{eqnarray}
Y &=& C_1 + C_2X+C_3X^2, \\
Y&\equiv&\log_{10}\left(T_{\rm coal}\over 10^{10}{\rm yr}\right), \\
X&\equiv&\log_{10}\left({M_1+M_2\over 10^6M_\odot}\right).
\end{eqnarray}
\label{eq:YvsX}
\end{subequations}
with
\begin{subequations}
\begin{eqnarray}
M_2/M_1=1:&&   C_1=-0.372,\ C_2=1.384,\ C_3=-0.025 \\
M_2/M_1=0.1:&& C_1=-0.478,\ C_2=1.357,\ C_3=-0.041.
\end{eqnarray}
\end{subequations}
Based on the figure, binary SBHs would be able to 
coalesce via interaction with stars alone in galaxies
with $M_\bullet \lesssim 2\times 10^6 M_\odot$.
For $M_1+M_2\gtrsim 10^7M_\odot$, evolution for 10 Gyr only
brings the binary separation slightly below $a_h$; in such
galaxies the most likely separation to find a massive binary
(in the absence of other sources of energy loss)
would  be near $a_h$.

The core continues to grow as the binary shrinks,
but the mass deficit is not related in a simple
way to  the mass in stars ``ejected'' by the binary
(e.g. \cite[Quinlan 1996]{Quinlan:96}).
Rather it results from a competition between
loss of stars to the binary, represented by $-{\cal F}(E,t)$, and 
the change in $N(E,t)$ due to diffusion of 
stars in energy, represented by
$-\partial F_E/\partial E$.
As the mass deficit increases, so do gradients
in $f$, which increases the flux of stars
toward the center and counteracts the drop in density.
In principle the two terms could balance,
but at some distance from the center the relaxation time
is so long that local $F_E(E)$ must drop below the
integrated loss term $\int_E^\infty {\cal F}(E) dE$ --
stars can not diffuse in fast enough to replace those
being lost to the binary and the density drops.
The Fokker-Planck solutions show that the mass deficit 
increases with binary binding energy as
\begin{equation}
M_{\rm def,c} \approx 1.7\left(M_1+M_2\right) \log_{10}\left(a_h/a\right)
\label{eq:mdefc}
\end{equation}
again with a weak dependence on $M_2/M_1$.
The mass deficit at the onset
of the gravitational radiation regime  is found to be
\begin{subequations}
\begin{eqnarray}
M_{\rm def,c} &\approx& (4.5,3.5,2.6,1.6)(M_1+M_2)\ \ (M_2/M_1=1) \\
            &\approx& (3.4,2.6,1.7,0.9)(M_1+M_2)\ \ (M_2/M_1=0.1)
\end{eqnarray}
\label{eq:mdefb}
\end{subequations}
for $M_1+M_2=(10^5,10^6,10^7,10^8)M_\odot$.
These values should be added to the mass deficits (\ref{eq:mdefh})
generated during formation of the binary when predicting
core sizes in real galaxies.

Are such mass deficits observed?
Only a handful of galaxies in the relevant mass range
($M_{\rm gal} \lesssim 10^{10} M_\odot$) are near enough
that their cores could be resolved even if present;
of these, neither the Milky Way nor M32 exhibit cores.
Also, as noted above, many low-luminosity spheroids have
compact central excesses rather than cores.
These facts do not rule out the past existence 
of massive binaries in these galaxies however.
(1) Binary evolution might have been driven more by
gas dynamical torques than by ejection of stars;
gas content during the most recent major merger is believed 
to be a steep inverse function of galaxy luminosity
 (\cite[Kauffmann \& Haehnelt 2000]{Kauffmann:00}).
(2) Star formation can create a dense core after the two 
SBHs have coalesced 
(\cite[Mihos \& Hernquist 1994]{Mihos:94}).
(3) A two-body relaxation time short enough to bring
the two SBHs together would also allow a Bahcall-Wolf
cusp to be regenerated in a comparable time after the two SBHs combine,
tending to erase the core (\cite[Merritt \& Szell 2006]{MS:06}).

From the point of view of physicists hoping to detect
gravitational waves,
it is disappointing that this model only guarantees coalescence
at the extreme low end of the SBH mass distribution.
(Astronomers hoping to detect binary SBHs may take
the opposite point of view.)
Fortunately, there is no dearth of ideas for
 overcoming the ``final parsec problem'' and allowing
binary SBHs to merge efficiently, even in massive galaxies:

\begin{figure}
  \vspace{0.2cm}
  \includegraphics[height=0.35\textheight]{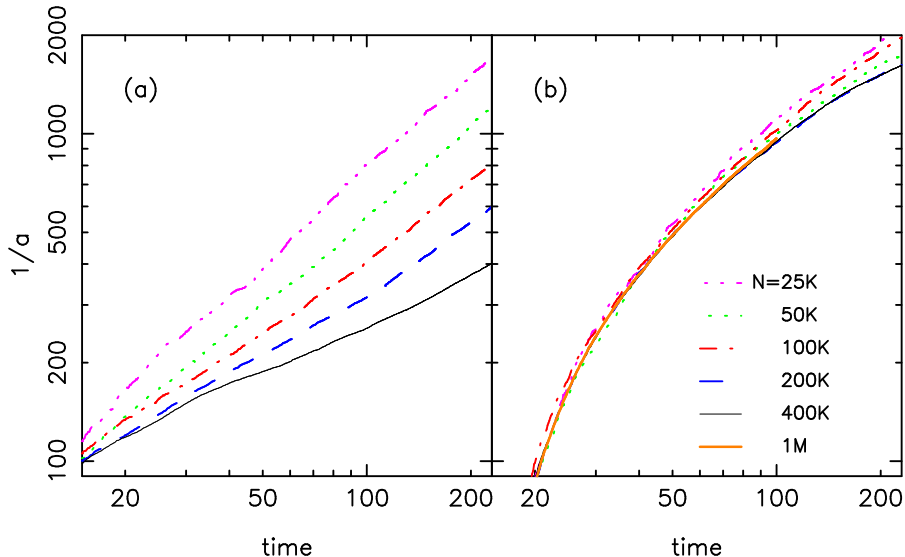}
  \caption{Efficient merger of binary SBHs in barred galaxies.
  Plots are based on $N$-body simulations (no gas) of
  equal-mass binaries at the centers of galaxy models,
  with and without rotation.
  {\it (a)} Spherical models.  The binary hardening rate
  declines with increasing $N$, as in figure~\ref{fig:MMS6},
  implying that the evolution
  would stall in the large-$N$ limit.
  {\it (b)} Binary evolution in a flattened, rotating version 
  of the same galaxy model.
  At $t\approx 10$, the rotating model forms a triaxial bar.
  Binary hardening rates in this model are essentially independent
  of $N$, indicating that the supply of stars to the binary
  is not limited by collisional loss-cone refilling as in 
  the spherical models.
  This is currently the only simulation that follows two SBHs 
  from kiloparsec to sub-parsec separations and that can be robustly
  scaled to real galaxies.
 (From Berczik {\it et al.} 2006.)}\label{fig:rotate}
\end{figure}

{\bf Non-axisymmetric geometries.} 
Real galaxies are not spherical nor even
axisymmetric; parsec-scale bars are relatively common
and departures from axisymmetry are often invoked 
to enhance fueling of AGN
(e.g. \cite[Shlosman {\it et~al.} 1990]{Shlosman:90}).
Orbits in a triaxial nucleus can be ``centrophilic,''
passing arbitrarily close to the center after a sufficiently long time
(Poon \& Merritt 2001, 2004).
This implies feeding rates for a central binary
that can approach  the ``full loss cone'' rate in 
spherical geometries, or 
\begin{equation}
{d\over dt} \left({1\over a}\right) 
\approx 2.5 F_{\rm c}{\sigma\over r_h^2}
\end{equation}
if the  fraction  $F_{\rm c}$ of centrophilic orbits is large
(\cite[Merritt \& Poon 2004]{Poon:04b}).
While $F_{\rm c}$ is impossible to know in any particular
galaxy, even small values imply much larger feeding rates
than in a diffusively-repopulated loss cone.
Figure~\ref{fig:rotate} shows results from $N$-body simulations
that support this idea.

{\bf Secondary slingshot.}
Stars ejected by a massive binary 
can interact with it again if they return to the nucleus
on nearly-radial orbits.
The total energy extracted from the binary via this 
``secondary slingshot'' will be the sum
of the discret energy changes during the interactions.
\cite{MM:03} showed that a mass ${\cal M}_\star$ of stars
initially in the binary's loss cone causes the binary to evolve as
\begin{equation}
{1\over a} \approx {1\over a_h} + {4\over r_h}\ln\left(1+{t\over t_0}\right),
\ \ \ \ t_0 = {2\mu\sigma^2\over {\cal M}_\star \langle\Delta E\rangle} P(E)
\end{equation}
in the absence of diffusive loss cone repopulation,
where $\langle\Delta E\rangle$ is the specific energy change after one 
interaction with the binary, $E$ is the initial energy
and $P(E)$ is the orbital period.
The secondary slingshot runs its course after a few orbital periods.
\cite{Sesana:07} sharpened this analysis by carrying out
detailed three-body scattering experiments and recording the
precise changes in energy of stars as they underwent repeated
interactions with the binary.
They inferred modest ($\sim \times 2$) changes in $1/a$ due
to the secondary slingshot,
but their assumption of a $\rho\sim r^{-2}$ density profile
around the binary was probably over-optimistic; such steep
density profiles are never observed and even if present 
initially would be rapidly destroyed when the binary first formed.

{\bf Bound subsystems.}
As noted above, recent observational studies have greatly
increased the number of galaxies believed to harbor compact 
nuclear star clusters; inferred masses for the clusters are 
comparable with the mass
that would normally be associated with a SBH.
It is not yet clear whether these subsystems co-exist with
SBHs, but if they do, they could provide an extra
source of stars to interact with a massive binary.
\cite[Zier (2006, 2007)]{Zier:06,Zier:07} explored this idea, assuming
a steeply rising density profile around the binary,
$\rho \propto r^{-\gamma}$, at the time that its
separation first reached $\sim a_h$. Zier concluded that
a cluster having total mass $\sim M_1+M_2$, distributed
as a steep power law, $\gamma\gtrsim 2.5$, could
extract enough energy from the binary to allow gravitational
wave coalescence in less than $10$ Gyr.
$N$-body tests of this hypothesis are sorely needed;
as in the \cite{Sesana:07} study, Zier's approach did
not allow him to self-consistently follow the effect of
formation of the binary on the surrounding mass distribution.

{\bf Masive perturbers.}
In a nucleus containing a spectrum of masses,
the gravitational scattering rate is proportional to
\begin{equation}
\tilde{m} = {\int n(m) m^2 dm \over \int n(m) m dm}
\end{equation}
(e.g. \cite[Merritt 2004]{Merritt:04}).
\cite{Perets:07a} argued that ``massive perturbers''
near the center of the Milky Way -- massive stars,
star clusters, giant molecular clouds -- are
sufficiently numerous to dominate $\tilde m$, implying
potentially much higher rates of gravitational scattering into a 
central sink than in the case of solar-mass perturbers.
\cite{Perets:07b} extended this argument to
galaxies in general, emphasizing in particular the early
stages following a galactic merger, 
and concluded that collisional loss cone repopulation
would be sufficient to guarantee
coalescence of binary SBHs in less than 10 Gyr for
all but the most massive binaries.
As in the studies of \cite{Sesana:07} and \cite{Zier:07},
\cite{Perets:07b} optimistically assumed a steep ($\rho\propto r^{-2}$)
density profile around the binary, in spite of $N$-body studies
showing rapid destruction of the cusps.
Their arguments for massive perturbers in giant E galaxies
are also rather speculative.

{\bf Multiple SBHs.}
An extreme case of a ``massive perturber'' is a third SBH,
which might scatter stars into a central binary 
(\cite[Zhao {\it et~al.} 2002]{Zhao:02}),
or perturb the binary directly, driving the two SBHs into
an eccentric orbit and shortening the time scale for gravitational 
wave losses (\cite[Valtonen {\it et~al.} 1994]{Valtonen:94};
\cite[Makino \& Ebisuzaki 1994]{Makino:94};
\cite[Blaes {\it et~al.} 2002]{Blaes:02}; 
\cite[Volonteri {\it et~al.} 2003]{Volonteri:03};  
\cite[Iwazawa {\it et~al.} 2006]{Iwasawa:06};
\cite[Hoffman \& Loeb 2007]{Hoffman:07}).
The likelihood of multiple-SBH systems forming is probably
highest in the brightest E galaxies since massive binaries
are most likely to stall (low stellar density, little gas) 
and since large galaxies experience the most frequent mergers.
Here again, more $N$-body simulations, including post-Newtonian
terms, are needed; among other
dynamical effects that could then be self-consistently included
are changes in core structure, and BH-core oscillations like those
described in the next section.

{\bf Gas.}
 The same galaxy mergers that create binary SBHs can 
also drive gas into the nucleus, and there is abundant
observational evidence for cold
(e.g. \cite[Jackson {\it et~al.} 1993]{Jackson:93};
\cite[Gallimore {\it et~al.} 2001]{Gallimore:01};
\cite[Greenhill {\it et~al.} 2003]{Greenhill:03})
and hot (e.g. \cite[Baganoff {\it et~al.} 2003]{Baganoff:03}) 
gas near the centers of at least some galaxies.
Dense concentrations of gas can substantially accelerate 
the evolution of a massive binary by increasing the drag on 
the individual BHs
(Escala {\it et~al.} 2004, 2005; \cite[Dotti {\it et~al.} 2007]{Dotti:07}).
The plausibility of such dense accumulations of gas,
with mass comparable to the mass of the SBHs, is unclear however
(e.g. \cite[Sakamoto {\it et~al.} 1999]{Sakamoto:99};
\cite[Christopher {\it et~al.} 2005]{Christopher:05}).
Large-scale galaxy merger simulations 
(\cite[Kazantzidis {\it et~al.} 2005]{Kazant:05};
\cite[Mayer {\it et~al.} 2007]{Mayer:07})
show that the presence of gas leads to more rapid formation
of the massive binary, but these simulations still lack the
resolution to follow the binary past $a\approx  r_h$
and so have nothing relevant to say (yet) concerning the 
final parsec problem.

\bigskip
As this summary indicates, many possible solutions
to the ``final parsec problem'' exist, but none is guaranteed to
be effective in all or even most galaxies.
The safest bet is that both coalesced and uncoalesced binary
SBHs exist, but with what relative frequency is still anyone's guess.

\section{SBH/IBH binaries}\label{IBH}

Secure dynamical evidence exists for SBHs in the mass range
$10^{6.5}\lesssim M_\bullet/M_\odot\lesssim 10^{9.5}$
(\cite[Ferrarese \& Ford 2005]{FF:05})
and compelling arguments have been made for BHs with masses
$10^5\lesssim M_\bullet/M_\odot\lesssim 10^7$
in active nuclei (\cite[Greene \& Ho 2004]{Greene:04}).
Binary mass ratios as extreme as $1000:1$, and possibly
greater, are therefore to be expected.
This possibility has received most attention in the context
of intermediate-mass black holes
(IBHs) in the Milky Way, where they could form 
in dense star clusters like the Arches or Quintuplet
before spiralling into the center and forming a tight
binary with the $\sim 3.5\times 10^6 M_\odot$ 
SBH (\cite[Portegies Zwart \& McMillan 2002]{Zwart:02};
\cite[Hansen \& Milosavljevi{\'c} 2003]{Hansen:03}).

\begin{figure}
  \vspace{0.5cm}
  \includegraphics[height=0.305\textheight]{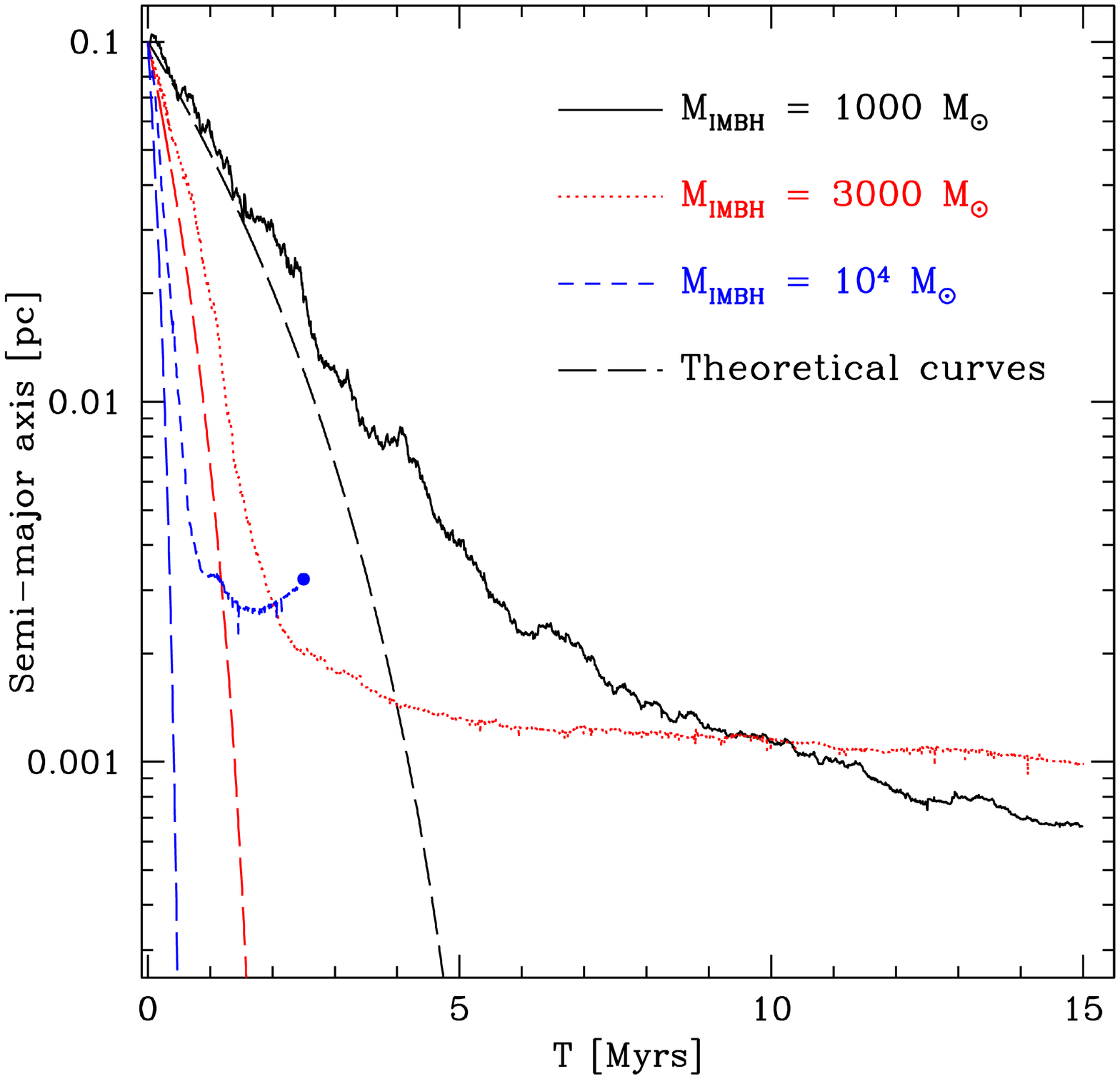}
  \includegraphics[height=0.31\textheight]{fig_RB1.ps}
  \caption{{\it Left:} $N$-body  simulations of the inspiral of IBHs into
  the center of the Milky Way. Solid lines show the separation
  between the two BHs and dashed lines are theoretical predictions
  that ignore loss-cone depletion or changes in the structure of the
  core.
  Smaller IBHs spiral in farther before ``stalling'';
  the $M_{\rm IBH}=10^3M_\odot$ simulation ends before the stalling
  radius is reached.
   (From Baumgardt, Gualandris \& Portegies Zwart 2006.)
   {\it Right:} Evolution beyond $a=a_h$, based on the Fokker-Planck
   model of Merritt, Mikkola \& Szell (2007). 
   Dashed lines indicate when the evolution time due to gravitational
   radiation losses is less than $10$ Gyr.}\label{fig:IBH}
\end{figure}

The predicted hard-binary separation for a SBH/IBH pair is
(\ref{eq:ah})
\begin{equation}
a_h \approx 0.5 {\rm mpc} \left({q\over 10^{-3}}\right) 
\left({M_\bullet\over 3.5\times 10^6M_\odot}\right)^{0.59},
\ \ \ \ q\equiv {M_{\rm IBH}\over M_\bullet}.
\end{equation}
This separation -- $\sim 10^2$ AU -- is comparable to
the periastron distances of the famous ``S'' stars
(\cite[Eckart {\it et~al.} 2002]{Eckart:02};
\cite[Ghez {\it et~al.} 2005]{Ghez:05}).
Dynamical constraints on the existence of an IBH at this
distance from the SBH are currently weak
(\cite[Yu \& Tremaine 2003]{Yu:03};
\cite[Hansen \& Milosavljevi{\'c} 2003]{Hansen:03};
\cite[Reid \& Brunthaler 2004]{Reid:04}).
Figure~\ref{fig:IBH}a shows $N$-body simulations designed to 
mimic inspiral of IBHs into the Galactic center.
The figure confirms the expected slowdown in the inspiral rate
at a separation $\sim a_h$.
Figure~\ref{fig:IBH}b plots evolutionary tracks for the same
three IBH masses as in the 
left panel, based on the Fokker-Planck model of
\cite{MMS:07}.
For $M_{\rm IBH}\lesssim 10^3 M_\odot$, evolution of the
binary is dominated by gravitational wave losses already
at $a=a_h$.

\begin{figure}
  \vspace{0.2cm}
  \includegraphics[height=0.315\textheight]{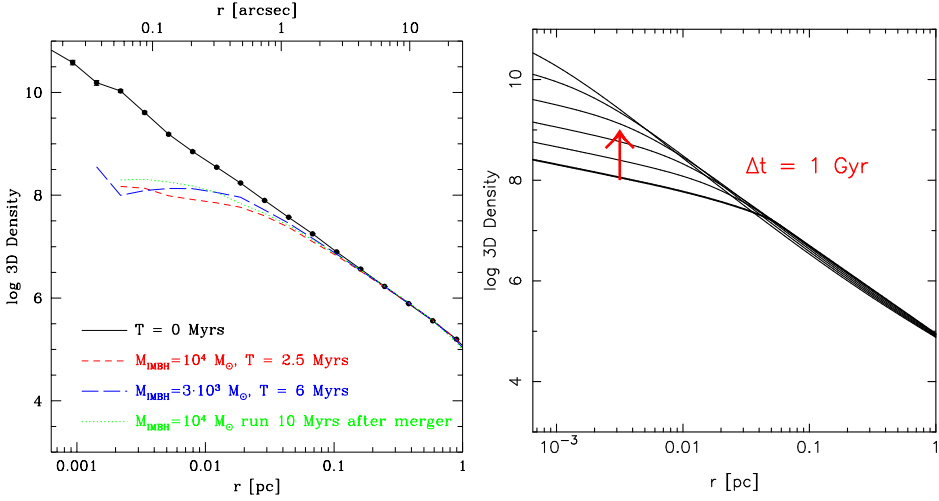}
  \includegraphics[height=0.335\textheight]{fig_baum_FP.ps}
  \caption{{\it Left:} Creation of a core by inspiral of IBHs
  into the Galactic center.  The initial density profile is shown
  by the solid line.  Green (dotted) line shows the core $10$ Myr 
  after the $10^4M_\odot$ IBH has merged with the SBH; almost
  no change occurs during this time.
   (From Baumgardt, Gualandris \& Portegies Zwart 2006.)
   {\it Right:} Fokker-Planck model showing how the cusp
   regenerates due to two-body scattering, on Gyr timescales.
   (Adapted from Merritt, Mikkola \& Szell 2007.)}\label{fig:IBH2}
\end{figure}

The same inward flux of stars that allows the binary to shrink
also implies an outward flux of stars ejected by the binary.
The latter are a potential source of ``hyper-velocity stars''
(HVSs),
stars moving in the halo with greater than Galactic escape velocity 
(\cite[Hills 1988]{Hills:98}).
The relation between the stellar ejection rate and the binary
hardening rate, when $a\le a_h$, 
is given by (\ref{eq:hardb}) after rewriting it as
\begin{equation}
T_{\rm hard}\equiv a{d\over dt}\left({1\over a}\right) \approx 
{2.5 \over M_\bullet} \times {\rm flux};
\label{eq:hardc}
\end{equation}
here ``flux'' is the total mass in stars per unit time,
from all energies, that are scattered
into (and ejected by) the binary.
Combining (\ref{eq:hardc}) with (\ref{eq:tdepend}), the flux is
\begin{subequations}
\begin{eqnarray}
&\sim& 5.0 {M_\bullet\over T_R(r_h)} 
\left[1+{5t\over T_R(r_h)}\right]^{-1/2} \\
&\lesssim& 350 M_\odot {\rm yr}^{-1} 
\end{eqnarray}
\end{subequations}
where the second line uses values appropriate to the Galactic center.
%This estimate is consistent with $N$-body results
%(\cite[Baumgardt, Gualandris \& Portegies Zwart 2006]{Baumgardt:06};
%\cite[Matsubayashi {\it et~al.} 2007]{Matsu:07}).
Relating the total ejected flux to the number of HVSs
that would be observed is not straighforward;
for instance, only a fraction ($\lesssim 10$\%) 
would be ejected with high enough velocity
to still be moving faster than $\sim 500$ km s$^{-1}$
after climbing through the Galactic potential
(\cite[Gualandris {\it et~al.} 2005]{Gualandris:05};
\cite[Baumgardt {\it et~al.} 2006]{Baumgardt:06}),
and targeted searches for HVSs only detect certain stellar types
so that knowledge of the stellar mass function is also required
(\cite[Brown {\it et~al.} 2006]{Brown:06}).

Inspiral of the IBH creates a core of radius 
$\sim 0.05\ {\rm pc} \approx 1''$ 
(figure~\ref{fig:IBH2}a).
Such a core might barely be detectable at the center of the 
Milky Way from star counts.
There is no clear indication of a core 
(\cite[Schoedel {\it et.~al} 2007]{Schoedel:07}),
but if the inspiral occurred more than a few Gyr
ago, star-star gravitational scattering would have
gone some way toward ``refilling'' the region depleted
by the binary 
(\cite[Merritt \& Wang 2005]{Wang:05};
\cite[Merritt \& Szell 2006]{MS:06}; figure~\ref{fig:IBH}b).
In this case, however, the ejected stars
would almost all have moved beyond the range of HVS surveys
by now.

The angular distribution of the ejected stars has been proposed
as a test for their origin; unlike other possible sources of
HVSs, a SBH/IBH binary tends to eject
stars parallel to the orbital plane or, if the orbit is eccentric,
in a particular direction 
(\cite[Levin 2006]{Levin:06};
\cite[Sesana {\it et~al.} 2006]{Sesana:06}).
In two $N$-body simulations of IBH inspiral however
(\cite[Baumgardt, Gualandris \& Portegies Zwart 2006]{Baumgardt:06};
\cite[Matsubayashi {\it et~al.} 2007]{Matsu:07}),
the orientation of the binary began to change appreciably,
in the manner of a random walk, after it became hard.
This was due to ``rotational Brownian motion''
(\cite[Merritt 2002]{Merritt:02}):
torques from passing stars -- the same stars that extract
energy and angular momentum from the binary -- also change
the direction of the binary's orbital angular momentum vector.
In one hardening time $|a/\dot{a}|$ of the binary,
its orientation changes by
\begin{subequations}
\begin{eqnarray}
\Delta\theta &\approx& q^{-1/2} \left({m_\star\over M_\bullet}\right)^{1/2} 
\left(1-e^2\right)^{-1/2} \\
&\approx& 9.0^\circ \left({q\over 10^{-3}}\right)^{-1/2} 
\left({M_\bullet\over 10^6 m_\star}\right)^{-1/2}
\left[{\left(1-e^2\right)^{-1/2}\over 5}\right].
\label{eq:RBM}
\end{eqnarray}
\end{subequations}
(The eccentricity dependence in (\ref{eq:RBM}) is approximate;
the numerical coefficient in this equation has only been confirmed by detailed
scattering experiments for $e=0$.)
In both of the cited $N$-body studies, the binary eccentricity
evolved appreciably away from zero before the orientation changes
became signficant.
Rotational Brownian motion  might not act quickly enough to randomize
the orienation of a SBH/IBH binary in a time of $\sim 10^8$ yr,
the flight time from the Galactic center to the halo,
unless perturbers more massive than Solar-mass stars
are present near the binary however
(\cite[Merritt 2002]{Merritt:02};
\cite[Perets \& Alexander 2007]{Perets:07}).

\section{Kicks and cores}\label{sec:kicksandcores}

After seeming to languish for several decades, the field
of numerical relativity has recently experienced exciting progress.
Following the breakthrough papers of \cite{Pretorius:05},
\cite{Campanelli:06} and \cite{Baker:06a}, several groups
have now successfully simulated the evolution of binary BHs 
all the way to coalescence.
The final inspiral is driven by emission of gravitational waves,
and in typical (asymmetric) inspirals, a net impulse -- a ``kick'' --
is imparted to the system due to anisotropic emission of the waves
(\cite[Bekenstein 1973]{Bekenstein:73}; \cite[Fitchett 1984]{Fitchett:84};
\cite[Favata et al. 2004]{Favata:04}).  Early arguments that the
magnitude of the recoil velocity would be modest for non-spinning BHs
%(\cite[Redmount \& Rees 1989]{Redmount:89}) 
were confirmed by the simulations,  which found
$V_{\rm kick}\lesssim 200$ km s$^{-1}$ in the absence of spins
(\cite[Baker et al. 2006]{Baker:06b}; 
\cite[Gonzalez et al. 2007a]{Gonzalez:07a};
\cite[Herrmann et al. 2007]{Herrmann:07}).  
The situation changed
dramatically following the first 
(\cite[Campanelli {\it et~al.} 2007a]{Campanelli:07a})
simulations of ``generic'' binaries,
in which the individual BHs were spinning and tilted with respect
to the orbital angular momentum vector.
Kicks as large as $\sim 2000$ km s$^{-1}$ have now
been confirmed (\cite[Campanelli {\it et~al.} 2007b]{Campanelli:07b}; 
\cite[Gonzalez {\it et~al.} 2007b]{Gonzalez:07b};
\cite[Tichy \& Marronetti 2007]{Tichy:07}), 
and scaling arguments based on the post-Newtonian approximation
suggest that the maximum kick velocity would
probably increase to $\sim 4000$ km s$^{-1}$ in the case of maximally-spinning
holes (\cite[Campanelli {\it et~al.} 2007b]{Campanelli:07b}).  
The most propitious configuration for the kicks appears to be
an equal-mass binary in which the individual
spin vectors are oppositely aligned and oriented parallel to the
orbital plane.  
For unequal-mass binaries, the maximum kick is
\begin{equation}
V_{\rm max} \approx 6\times 10^4 {\rm km\ s}^{-1} {q^2\over (1+q)^4}
\end{equation}
where $q\equiv M_2/M_1\le 1$ is the binary mass ratio
and maximal spins have been assumed
(\cite[Campanelli {\it et~al.} 2007c]{Campanelli:07c}).
Orienting the BHs with their spins perpendicular to the orbital
angular momentum may seem odd
(\cite[Bogdanovi\'c, Reynolds \& Miller 2007]{BRM:07}),
but there is considerable evidence that SBH spins bear no relation
to the orientations of the gas disks that surround them
(e.g. \cite[Kinney {\it et~al.} 2000]{Kinney:00}; \cite[Gallimore {\it et.~al.} 2006]{Gallimore:06}) and this is presumably even more true with
respect to the directions of infalling BHs.
Galaxy escape velocities are $\lesssim 3000$ km s$^{-1}$ 
(\cite[Merritt {\it et~al.} 2004]{MMFHH:04}),
so gravitational wave recoil can in principle eject 
coalescing SBHs completely from galaxies.

\begin{figure}
 \vspace{-0.2cm}
 \includegraphics[height=14cm]{fig_cont_1.ps}
 \includegraphics[height=15.cm]{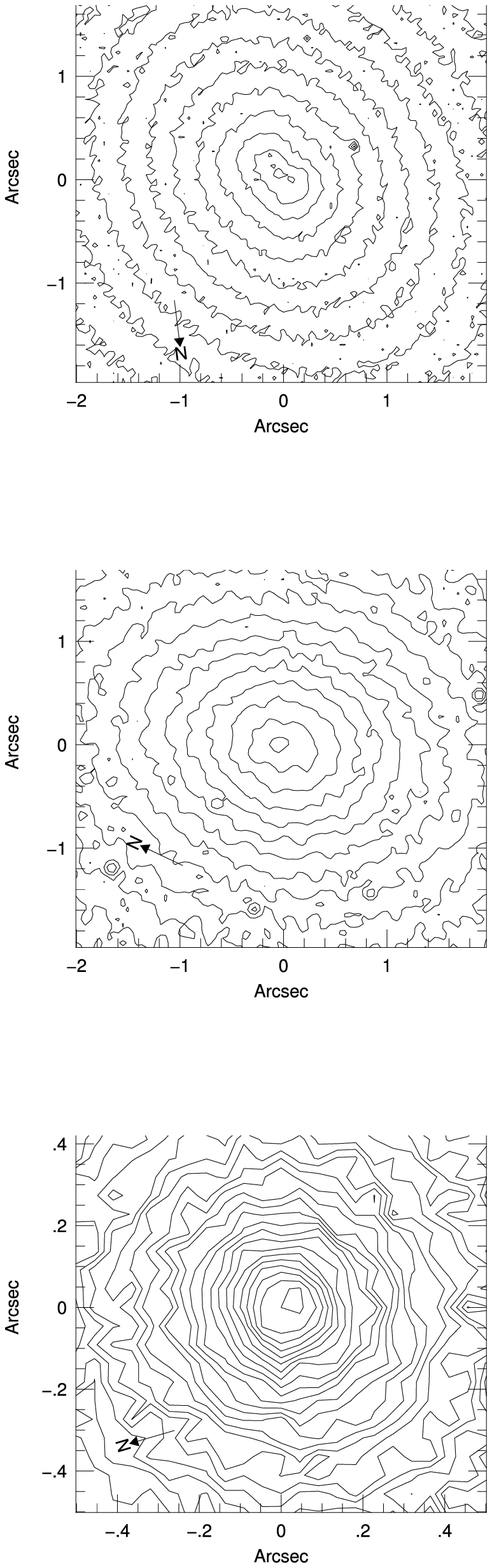}
 \caption{{\it Left:} Core oscillations in an $N$-body simulation
  of ejection of a SBH from the center of a galaxy; 
  the kick velocity was 60\% of the escape velocity.  
  Contour plots show the stellar density at equally spaced times,
  spanning $\sim 1/2$ of the SBH's orbital period.
  Filled circles
  are the SBH and crosses indicate the location of the (projected)
  density maxima. (From Gualandris \& Merritt 2007.)
  {\it Right:} Surface brightness contours of three ``core'' galaxies 
  with double or offset nuclei, from Lauer {\it et~al.} (2005).
  {\it Top:} NGC 4382; {\it middle:} NGC 507; {\it bottom:} NGC 1374.
  \label{fig:cont}}
\end{figure}

\begin{figure}
  \vspace{0.5cm}
  \includegraphics[height=0.62\textheight,angle=-90.]{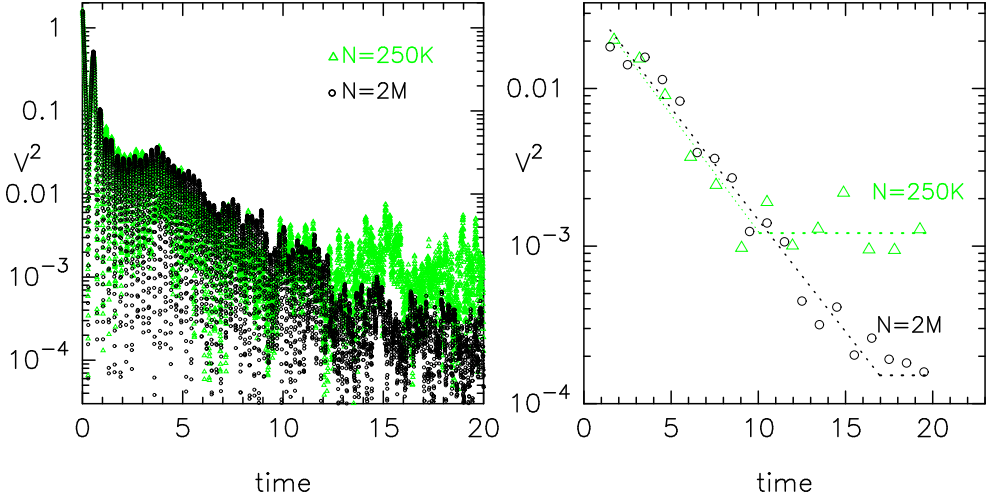}
  \caption{Evolution of the SBH kinetic energy following a kick of
    60\% the central escape velocity, in two $N$-body simulations of
    a galaxy represented by $N$ stars, with
    $N=(2.5\times 10^5, 2\times 10^6)$.
    The mass of the SBH, and the total mass of the galaxy, 
    are the same in the two simulations; all that varies 
    is the mass of the ``star'' particles.
    The right-hand panel shows binned values of $V^2$.
    Most of the elapsed time is spent in SBH/core oscillations
    like those illustrated in figure~\ref{fig:cont}.
    Eventually, the SBH's kinetic energy decays to the Brownian
    value, shown as the horizontal dashed lines in the right panel.
    The Brownian velocity scales as $m_\star^{1/2}$ and so is
    smaller for larger $N$.
    Scaled to a $3\times 10^{10}M_\odot$ galaxy, the
    time to reach the Brownian regime would be $\sim 10^8$ yr.
    (Adapted from Gualandris \& Merritt 2007.)
   }\label{fig:tests}
\end{figure}

Detailed $N$-body simulations show that the motion of a SBH that 
has been kicked with enough velocity to eject it out of the 
core, but not fast enough to escape the galaxy entirely, 
exhibits three distinct phases (\cite[Gualandris \& Merritt 2007]{GM:07}):

\begin{itemize}
\item {\it Phase I:}
The SBH oscillates with decreasing amplitude, losing
  energy via dynamical friction each time it passes through the core.
  Chandrasekhar's theory accurately reproduces the motion of the SBH
  in this regime for values $2\lesssim\ln\Lambda\lesssim 3$ of the
  Coulomb logarithm, if the gradually-decreasing core density 
  is taken into account.  
\item {\it Phase II:} When the amplitude of the motion has decayed
   to roughly the core radius, the SBH
  and core begin to exhibit oscillations about their common center of
  mass (figure~\ref{fig:cont}). These oscillations decay exponentially
  (figure~\ref{fig:tests}), but with a time constant that is $10-20$ times 
  longer than would be predicted by a naive application of the 
  dynamical friction formula.
\item {\it Phase III:} Eventually the SBH's kinetic energy drops
  to an average value
\begin{equation}
{1\over 2}M_\bullet V_\bullet^2 \approx {1\over 2}m_\star {\rm v}_\star^2
\label{eq:vbrown}
\end{equation}
i.e. to the kinetic energy of a single star.
This is the regime of gravitational Brownian motion
(\cite[Bahcall \& Wolf 1976]{BW:76}; \cite[Young 1977]{Young:77}; 
\cite[Merritt {\it et~al.} 2007]{MBL:07}).
\end{itemize}

A natural definition of the ``return time'' of a kicked SBH is the 
time to reach the Brownian regime.  
Unless the kick is very close to the escape velocity,
the return time is dominated by the time spent in ``Phase II;''
during this time, the SBH's energy decays roughly as 
\begin{equation}
E \approx \Phi(0) + \Phi(r_c) e^{-(t-t_c)/\tau}
\end{equation}
(\cite[Gualandris \& Merritt 2007]{GM:07});
$t_c$ is the time when the SBH re-enters the core whose
radius is $r_c$.
The damping time in the $N$-body simulations, $\tau$, is 
\begin{subequations}
\begin{eqnarray}
\tau &\approx & 15{\sigma_c^3\over G^2\rho_cM_\bullet}\\
&\approx &1.2\times 10^7{\rm yr} 
\left({\sigma_c\over 250{\rm km\ s}^{-1}}\right)^3
\left({\rho_c\over 10^3M_\odot {\rm pc}^{-3}}\right)^{-1}
\left({M_\bullet\over 10^9M_\odot}\right)^{-1}, 
\end{eqnarray}
\end{subequations}
with $\rho_c$ and $\sigma_c$ the core density and velocity
dispersion respectively.
The number of decay times required for the SBH's energy
to reach the Brownian level is 
$\sim\ln\left({M_\bullet/m_\star}\right)
\approx 20$, implying that a kicked SBH will remain significantly
off-center for a long time, as long as $\sim 1$ Gyr in a bright galaxy
with a low-density core.

In fact, asymmetric cores are rather common.
These include off-center nuclei 
(\cite[Bingelli {\it et~al.} 2000]{Bingelli:00};
\cite[Lauer {\it et~al.} 2005]{Lauer:05});
double nuclei (\cite[Lauer {\it et~al.} 1996]{Lauer:96});
and cores with a central minimum in the surface brightness
(\cite[Lauer {\it et~al.} 2002]{Lauer:02}).
Three examples, from \cite{Lauer:05}, are reproduced here
on the right side of figure~\ref{fig:cont}; 
all are luminous ``core'' galaxies, and each
strikingly resembles at least one frame from the $N$-body montage
on the left.
The longevity of the ``Phase II'' oscillations makes the kicks
a plausible model for the observed asymmetries.
This explanation is probably not appropriate for the famous
double nucleus of M31, since M31 is not a ``core'' galaxy,
and since one of the brightness peaks in M31 (the one associated
with the SBH) lies essentially at the galaxy photocenter;
Figure~\ref{fig:cont} suggests that an oscillating SBH
would typically (though not always) be found on the opposite side 
of the galaxy from the point of peak brightness.
The M31 double nucleus has been successfully modelled as a clump
of stars on eccentric orbits which maintain their lopsidedness
by virtue of moving deep within the Keplerian potential of the SBH
(\cite[Tremaine 1995]{Tremaine:95}).

\begin{figure}
 \vspace{0.2cm}
 \includegraphics[height=0.30\textheight]{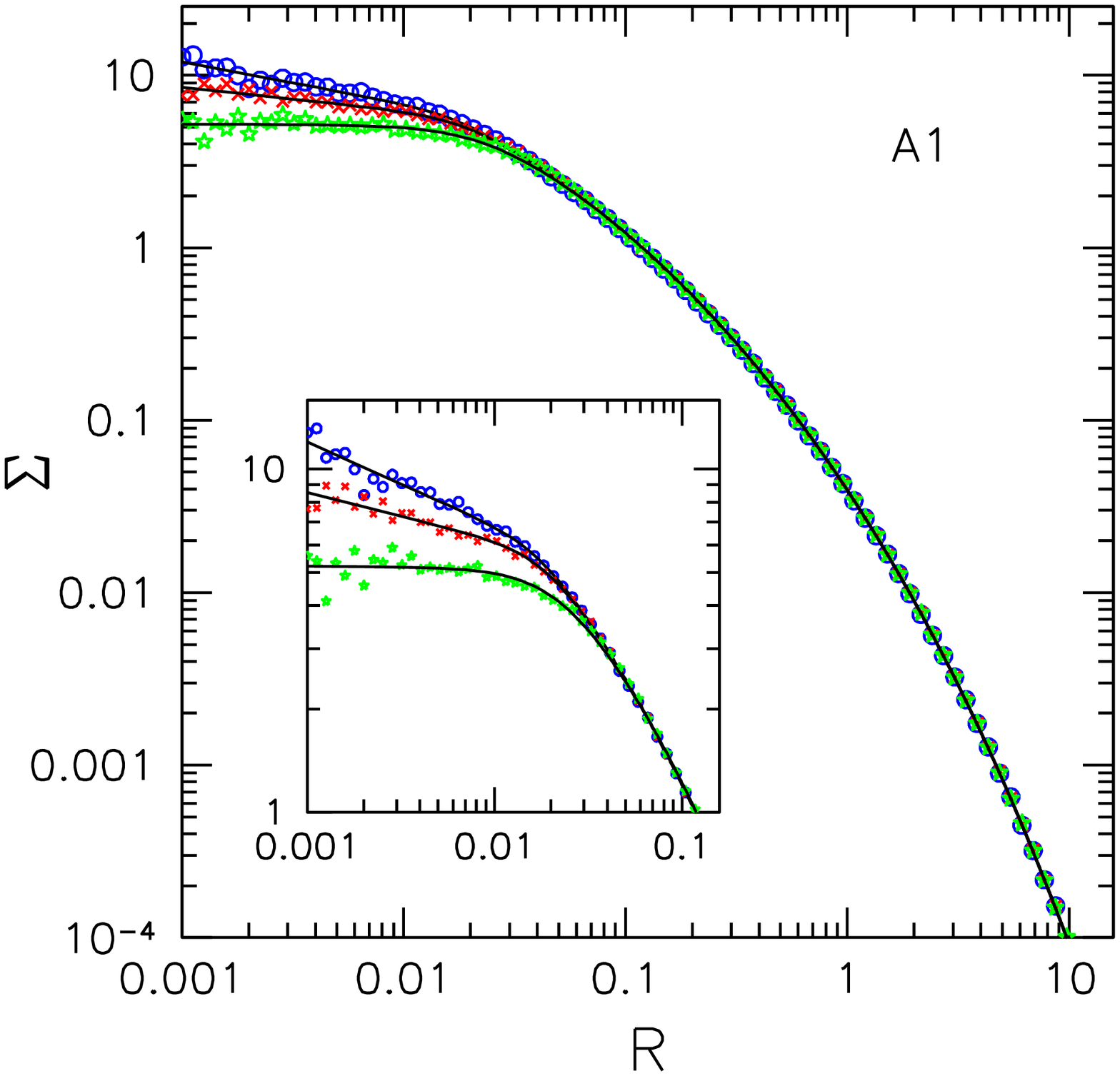}
 \includegraphics[height=0.30\textheight]{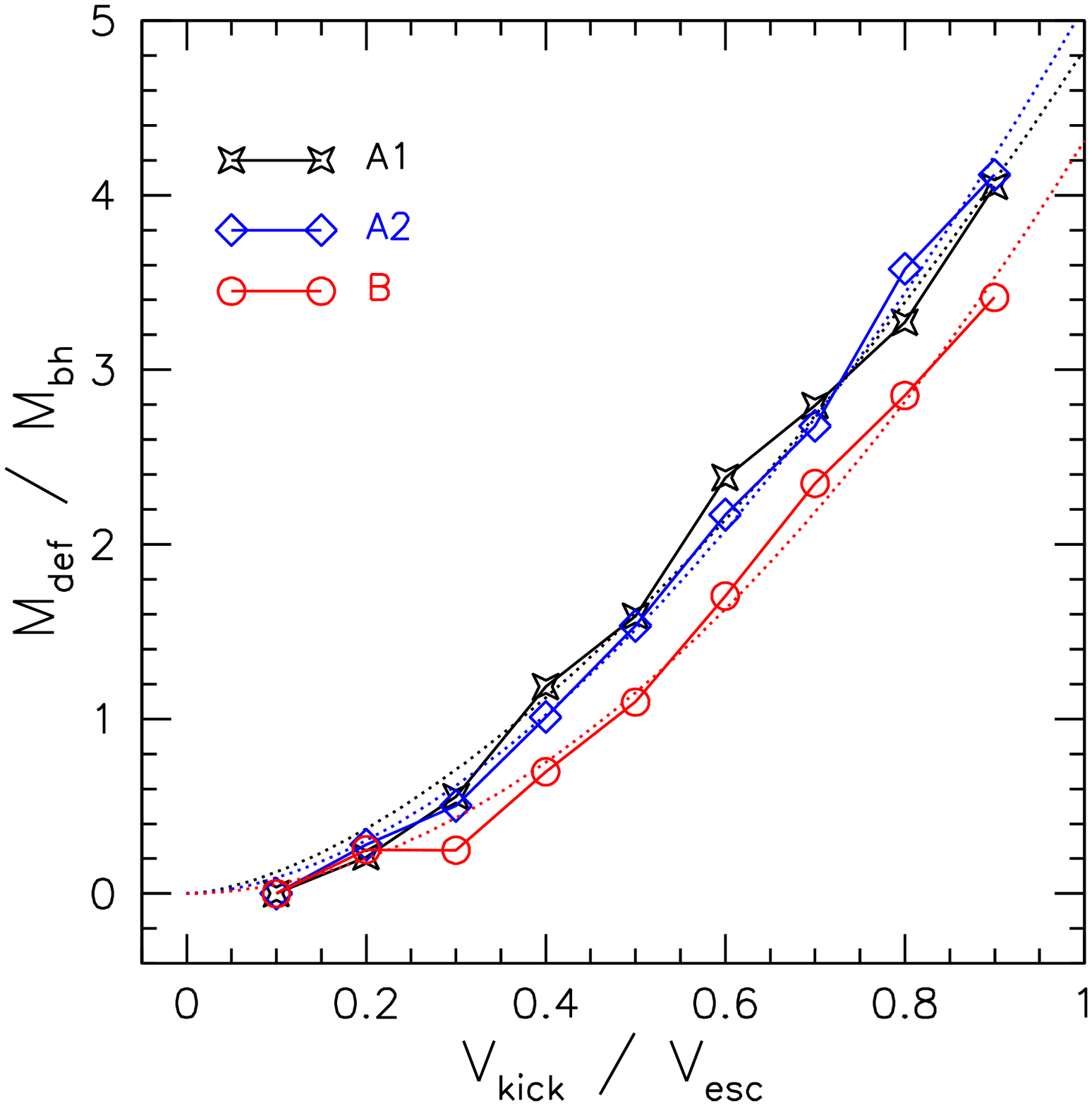}
 \caption{{\it Left:} 
  Points show projected  density profiles computed from $N$-body
  models after a kicked SBH has returned to the center,
  for three different values of the kick velocity 
  ($V_{\rm kick}=(0.2,0.4,0.8)\times V_{\rm esc}$). 
  Each set of points is compared with the best-fitting core-Sersic
  model (lines). 
  The insert shows a  zoom into the central region.
  {\it Right:}  Mass deficits generated by kicked SBHs.
  The differents sets of points correspond to different
  galaxy models.
  (Adapted from Gualandris \& Merritt 2007.)
  \label{fig:GMmdef}}
\end{figure}

The kicks are quite effective at inflating cores
(\cite[Merritt {\it et~al.} 2004]{MMFHH:04};
\cite[Boylan-Kinchin {\it et~al.} 2004]{Boylan:04}).
Figure~\ref{fig:GMmdef}, from \cite{GM:07}, illustrates this:
the mass deficit generated by the kick is approximately
\begin{equation}
M_{\rm def,k} \approx 5M_\bullet \left(V_{\rm kick}/V_{\rm esc}\right)^{1.75}.
\label{eq:mdefk}
\end{equation}
Even modest kicks can generate cores substantially larger 
than those produced during the formation of a massive binary
($\sim 1 M_\bullet$; eq.~\ref{eq:mdefh}).
Furthermore, this mechanism is potentially effective in even the
most luminous galaxies, unlike the relaxation-driven model
for core growth discussed above (eq.~\ref{eq:mdefc})
which only applies to nuclei with short relaxation times.
Gravitational radiation recoil is therefore a tenable explanation 
for the subset
of luminous E galaxies with large mass deficits (figure~\ref{fig:mdef}). 
An alternative explanation for the over-sized cores
(\cite[Lauer {\it et~al.} 2007]{Lauer:07})
postulates that the SBHs in these galaxies are ``hypermassive,'' 
$M_\bullet\gtrsim 10^{10}M_\odot$ and that the cores
are a consequence of slingshot ejection by a massive binary.

The kicks have a number of other potentially observable 
consequences, including spatially and/or kinematically offset AGN
(\cite[Madau \& Quataert 2004]{MQ:04}; 
\cite[Haehnelt {\it et~al.} 2006]{Haehnelt:06};
\cite[Merritt {\it et~al.} 2006]{Naked:06};
\cite[Bonning {\it et~al.} 2007]{Bonning:07})
and distorted or wiggling radio jets
(\cite[Gualandris \& Merritt 2007]{GM:07}).
Many of these manifestations were first discussed by R. Kapoor in
a remarkably prescient series of papers
(\cite[Kapoor 1976,1983a,b,1985]{Kapoor:76,Kapoor:1983a,Kapoor:1983b,
Kapoor:85}).

\section{Black-hole-driven expansion}

The growth of a core around a shrinking, binary SBH
was discussed above: beyond a certain radius, the
relaxation time becomes so long that the encounter-driven 
flux of stars toward the center cannot compensate for 
losses to the binary, forcing the density to drop.
A similar process takes place around a single SBH
(\cite[Shapiro 1977]{Shapiro:77}; \cite[Dokuchaev 1989]{Dokuchaev:89}): 
stars coming too near are consumed, or disrupted, and the
density drops.
This effect is absent from the classical equilibrium
models for stars around a BH
(e.g. \cite[Bahcall \& Wolf 1976]{BW:76}; \cite[Cohn \& Kulsrud 1978]{CK:78})
since these solutions fix the phase-space density far from the BH,
enforcing an inward flux of stars 
precisely large enough to replace the stars 
being consumed by the sink.
In reality, the BH acts as a heat source, in much the same way that 
hard binary stars inject energy into a post-core-collapse globular 
cluster and cause it to re-expand.

\begin{figure}
\vspace{-2.cm}
\begin{center}
\includegraphics[width=0.65\textwidth,angle=-90.]{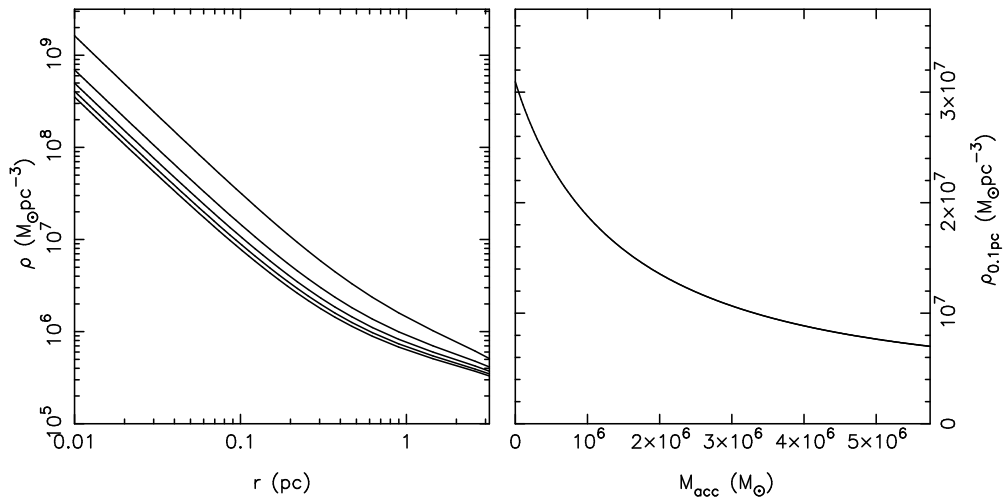}
\caption
{Black-hole-driven expansion of a nucleus; this Fokker-Planck
model was given parameters such that the density at the end
is similar to what is currently observed in the nucleus
of M32, with a final influence radius $r_h\approx 3$ pc.
The left panel shows density profiles at constant time intervals
after a Bahcall-Wolf cusp has been established;
the right panel shows the evolution of the density at
$0.1$ pc as a function of $M_{acc}$, 
the accumulated mass in tidally-disrupted stars.
As scaled to M32, the final time is roughly $2\times 10^{10}$ yr.
This plot suggests that the densities of collisional
nuclei like those of M32 and the Milky Way were once higher, 
by factors of $\sim $ a few, than at present.
(From Merritt 2006b.)
\label{fig:m32expand}
}
\end{center}
\end{figure}

A simple model that produces self-similar
expansion of a nucleus containing a SBH can be constructed
by simply changing the outer boundary condition in the
\cite{BW:76} problem
from $f(0)=f_0$ to $f(0)=0$.
One finds that the evolution after $\sim$one relaxation
time can be described as 
$\rho(r,t) = \rho_c(t)\rho^*(r)$, with
$\rho^*(r)$ slightly steeper than the $\rho\sim r^{-7/4}$
Bahcall-Wolf form; the normalization drops
off as $\rho_c\propto t^{-1}$ at late times.
Figure~\ref{fig:m32expand} shows the results of
a slightly more realistic calculation in a model
designed to mimic the nearby dE galaxy M32.
After reaching approximately the Bahcall-Wolf form, 
the density drops in amplitude
with roughly fixed slope for $r\lesssim r_h$.
This example suggests that the nuclei of galaxies
like M32 or the Milky Way might have been $\sim$ a few
times denser in the past than they are now,
with correspondingly higher rates of stellar
tidal disruption and stellar collisions.

Expansion due to a central BH has been observed in a 
handful of studies based on fluid 
(\cite[Amaro-Seoane {\it et~al.} 2004]{Amaro:04}),
Monte-Carlo (\cite[Shapiro \& Marchant 1978]{SM:78};
\cite[Marchant \& Shapiro 1980]{MS:80};
\cite[Freitag {\it et~al.} 2006]{Freitag:06}),
Fokker-Planck (\cite[Murphy {\it et~al.} 1991]{MCD:91}),
and $N$-body (\cite[Baumgardt {\it et~al.} 2000]{Baumgardt:04}) algorithms.
All of these studies allowed stars to be lost into or destroyed
by the BH; however most adopted parameters more
suited to globular clusters than to nuclei,
e.g. a constant-density core.
\cite{MCD:91} applied the isotropic, multi-mass
Fokker-Planck equation to the evolution of nuclei
containing SBHs, including an approximate loss term
in the form of (\ref{eq:dndt}) to model the
scattering of low-angular-momentum stars into the SBH.
Most of their models had what would now be considered
unphysically high densities and the evolution was dominated
by physical collisions between stars.
However in two models with lower densities, they reported
observing significant expansion over $10^{10}$ yr; 
these models had initial central relaxation times of 
$T_r\lesssim 10^9$ yr when scaled to real galaxies,
similar to the relaxation times near the centers of
M32 and the Milky Way.
The $\rho\sim r^{-7/4}$ form of the density profile
near the SBH was observed to be approximately conserved
during the expansion.
\cite{Freitag:06} carried out Monte-Carlo
evolutionary calculations of a suite of models
containing a mass spectrum, some of which were
designed to mimic the Galactic center star cluster.
After the stellar-mass BHs in their models
had segregated to the center,
they observed a roughly self-similar expansion.
\cite{Baumgardt:04} followed core collapse
in $N$-body models with and without a 
massive central particle;
``tidal destruction'' was modelled by simply
removing stars that came within a certain distance
of the massive particle.
When the ``black hole'' was present, the cluster 
expanded almost from the start
and in an approximately self-similar way.
These important studies notwithstanding, there is a crucial
need for more work on this problem in order to
understand how the rates of processes like stellar tidal
disruption vary over cosmological times
(e.g. \cite[Milosavljevi{\'c} {\it et~al.} 2006]{milos:06}).

\begin{acknowledgments}
We thank H. Baumgardt, A. Graham and T. Lauer for permission to
reproduce figures from their published work, and A. Graham
for comments on the manuscript.
We acknowledge support from the
National Science Foundation under grants 
AST-0420920 and AST-0437519 and from the National
Aeronautics and Space Administration under grants
NNG04GJ48G and NNX07AH15G.
\end{acknowledgments}

\end{document}